\documentclass[useAMS,usenatbib,letterpaper]{mn2e} 
\usepackage[bookmarks]{hyperref}
\usepackage{graphicx}
\usepackage[T1]{fontenc}
\usepackage{ae,aecompl}
\pdfoutput=1 
\usepackage[totalwidth=500pt,totalheight=680pt]{geometry}
\newcommand{\be}{\begin{equation}}
\newcommand{\ee}{\end{equation}} 
\newcommand{\bse}{\begin{subequations}}
\newcommand{\ese}{\end{subequations}} 
\newcommand{\bary}{\begin{eqnarray}}
\newcommand{\eary}{\end{eqnarray}} 
 
\newcommand{\epsf}{ \epsilon_{\rm f}}
\newcommand{\epsr}{\epsilon_{\rm r}}
\newcommand{\mbh}{M_{\rm BH}}

\newcommand{\mdotbh}{\dot{m}_{\rm BH}}

\newcommand{\msun}{{\rm M}_{\odot}}
\newcommand{\mseed}{ m_{\rm seed}} 
\newcommand{\rcirc}{r_{\rm circ}}
\newcommand{\rbondi}{ r_{\rm B}} 
\newcommand{\tvisc}{ t_{\rm visc}}
\newcommand{\tbondi}{ t_{\rm Bondi}} 
\newcommand{\mdotbondi}{ \dot{m}_{\rm Bondi}} 
\newcommand{\mstar}{ {M}_{\rm star}}
\newcommand{\K}{\rm K}
\newcommand{\Vphi}{ V_{\rm \phi}}
\newcommand{\cs}{  c_{\rm s}}
\newcommand{\Cvisc}{ C_{\rm visc}}
\newcommand{\mcrit}{ M_{200}}

\newcommand{\lsim}{\mathrel{\hbox{\rlap{\lower.55ex\hbox{$\sim$}} \kern-.3em\raise.4ex\hbox{$<$}}}}
\newcommand{\gsim}{\mathrel{\hbox{\rlap{\lower.55ex\hbox{$\sim$}} \kern-.3em\raise.4ex\hbox{$>$}}}}
\newcommand{\Msolar}{\,\rm M_{\odot}}

\def\kms{ \rm km\,s^{-1}}
\def\pc{\rm pc}
\def\kpc{\rm kpc}
\def\Mpc{\rm Mpc }
\def\Myr{\rm Myr}
\def\yr{\rm yr}
\def\Gyr{\rm Gyr}

\title[The impact of angular momentum on black hole accretion]{The impact of 
angular momentum on black hole accretion rates in simulations of galaxy formation}
\author[Y. M. Rosas-Guevara et al. ]{Y. M. Rosas-Guevara $^{1}$ \thanks{email:
y.m.rosas-guevara@dur.ac.uk (YRG)}, R. G. Bower $^{1}$ \thanks{email:r.g.bower@dur.ac.uk (RGB)}, 
J. Schaye $^2$, M. Furlong $^{1}$, C. S. Frenk $^{1}$,
\newauthor
C. M. Booth $^{3}$, R. A. Crain $^{2,4}$, C. Dalla Vecchia $^{5}$, M. Schaller $^{1}$, 
T. Theuns $^{1,6}$.\\ 
$^{1}$ Institute for Computational
Cosmology (ICC), Department of Physics, University of Durham, South Road, Durham, DH1 3LE, UK.
\\
$^{2}$ Leiden Observatory, Leiden University, P.O. Box 9513, 2300 RA Leiden, The Netherlands.
\\
$^{3}$ Department of Astronomy and Astrophysics, The University of Chicago, Chicago, IL 60637.
\\
$^{4}$ Astrophysics Research Institute, Liverpool John Moores University, 146 Brownlow Hill, Liverpool, L3 5RF
\\
$^{5}$ Max Planck Institute for Extraterrestrial Physics, Gissenbachstra\ss{}e 1, 85748 Garching, Germany.
\\
$^{6}$ Department of Physics, University of Antwerp, Campus Groenenborger, Groenenborgerlaan 171, B-2020 Antwerp, Belgium.
}
\begin{document}

\date{\today}

\pagerange{\pageref{firstpage}--\pageref{lastpage}} \pubyear{2013}

\maketitle

\label{firstpage}

\begin{abstract}
Feedback from energy liberated by gas accretion onto black holes
(BHs) is an attractive mechanism to explain the exponential cut-off
at the massive end of the galaxy stellar mass function (SMF). Most previous
implementations of BH accretion in hydrodynamical simulations
of galaxy formation have assumed that BHs grow at an accretion rate
that is proportion to the Bondi rate. A major concern is that the Bondi accretion
  rate is inappropriate when the accreting material has significant
  angular momentum.  We present an improved accretion model that takes
  into account the circularisation and subsequent viscous transport of
  infalling material, and implemented as a ``subgrid'' model in
  hydrodynamic simulations.  The resulting accretion rates are generally low in low
  mass ($\lsim 10^{11.5} \msun$) halos, but show outbursts of
  Eddington-limited accretion during galaxy mergers. During outbursts
  these objects strongly resemble quasars. In higher mass haloes, gas
  accretion peaks at $\sim10$\% of the Eddington rate, which is thought to be conducive to the formation of radio jets.  
The resulting accretion rate depends strongly on the effective pressure of the gas surrounding
the BH, which in turn depends strongly on halo mass.  This induces a sharp transition in the importance of BH feedback.
In small haloes,the growth of galaxies is regulated by star formation and supernova
feedback, but above a halo mass of $10^{11.5}\msun$, rapid BH growth leads to the 
suppression of star formation and reduced growth of stellar mass with increasing halo mass.
\end{abstract}
 

\section{Introduction}  
 
A fundamental open question in galaxy formation is the role that black
holes play in shaping the galaxy around them.  The observed scaling relations
between the mass of the central supermassive black hole (BH) and the properties of
the bulge \citep{magorrian1997,tremaine2002,mullaney2012} suggest that
there is an intimate connection between the growth of the central BH and 
the evolution of its host galaxy. From these observations, however, it is not clear 
whether the formation of the BH plays an integral part in the galaxy formation 
process, or whether it is simply a bi-product of the process of galaxy's evolution. There sure two lines
of argument that suggest that the first option is correct. Firstly, the energies
that are available from the formation of a $10^9 \msun$ BH are enormous,
greatly exceeding the binding energy of a galaxy's baryonic halo. Unless the
coupling of the accretion energy to the surrounding baryons is extremely weak,
it would be surprising if the formation of the BH has little
impact on its surroundings. Secondly, there is a strong observed correlation
between the mechanical power of radio galaxy lobes in galaxy clusters and the
cluster gas cooling rate.  Many authors have argued that the energy deposited by
the radio galaxy is sufficient to replenish the cooling radiation of the system
\citep{mcNamara2007}. Careful observations of
galaxy groups have revealed evidence of similar levels of energy input to radio
galaxies in groups \citep{antognini12,birzan2012,ma2013}; 
this regime is more relevant to the connection between BH and galaxy growth.

In phenomenological or semi-analytic models, feedback from 
AGNs is an indispensable element that enables the models to reproduce  the stellar mass
function of the local universe \citep{croton2006,bower2006,bower2008}.
AGN feedback is assumed to be ineffective in low mass haloes, where
the gas cooling time is short compared to the sound-crossing time
\citep{white_frenk91}, and  only to couple effectively in quasi-hydrostatic haloes
($M\gsim 10^{12}\msun$). 
This dichotomy has some observational support, since the bulk of energy output 
from quasars
appears to be radiated, while the mechanical energy of radio galaxies is
trapped in the overall potential. As a  result, a characteristic mass scale is
introduced  where the SMF presents a break: accretion in low mass haloes is
dominated by cold and rapidly cooling gas (since the cooling time is less than
the free-fall or sound-crossing time), while accretion in high mass halo occurs
through quasi-hydrostatic cooling gas flows (where the gas is approximately in
pressure balance, and the sound crossing time is less than the cooling time) \citep{white_frenk91}.
The importance of this distinction can be understood if the primary driver of
the star formation rate in galaxies is the balance between outflows and inflows
(i.e. the star formation rate of galaxies adjusts itself so that the inflow and
outflow are in equilibrium). In the case of low mass haloes, the AGN feedback
loop is (assumed to be) ineffective and the balance between gas supply and
outflow is set by the supernova driven outflow rate.  In higher mass haloes,
the AGNs regulate the galaxy growth either by offsetting the cooling rate
\citep{bower2006}, or by puffing up the hot gas halo
\citep{bower2008,McCarthy2011,bower2012} so that the cooling rate is reduced. In either
case, the result is to suppress the mass of the cold gas and reduce the star
formation rate in massive haloes, creating a break in the stellar mass function. 

In this scenario, the distinction between rapid cooling and hydrostatic haloes
is critical.  In the absence of a clear physical scale at which BH feedback
becomes effective, the stellar mass function behaves as a power--law
\citep{benson2003,bower2012} because the impact of gas ejection builds up over
a wide range of stellar mass. However, semi-analytic models make a variety of
simplifying assumptions, and it is possible that AGN-driven and star-formation
driven outflows might not combine as simply as is envisaged.

Hydrodynamic  simulations have the great advantage that there is no need to
make an explicit distinction between hydrostatic and rapidly cooling haloes.
Any dependence on the ratio of cooling and dynamical timescales should emerge
from the solution of the hydrodynamic equations. There is a long history of
papers that include AGN feedback in numerical simulations. \cite{dimatteo2005} and \cite{springel2005} 
introduced BH fuelling in order to 
study the impact of BH accretion on galaxy mergers \citep{hopkins2006,hopkins2007,sijacki2007,barai2013}.  
More recently, some papers have focused on modelling  both feedback modes, 
{\it quasar} and {\it radio}, either making use of two accretion modes  \citep{sijacki2007,vogelsberger2014} or 
different ways of implementing jet feedback \citep{dubois2010,debuhr2011}. 

The approach we follow here is extensively based on the development of the
\cite{springel2005} accretion model presented in \cite{booth_schaye09} (hereafter BS09). 
\cite{booth_schaye10} emphasise the importance of self regulation: BHs grow until their
energy output is comparable to the binding energy of the halo, resulting in a
very tight correlation between halo mass and BH mass.
They demonstrate that the slope of the correlation departs
from unity because of the variation of halo concentration, and that changes in
the feedback efficiency result in an offset in BH mass such that the
rate by which the BH releases energy remains fixed. These simulations
focused on the physics of BH accretion at high masses, and therefore
used a relatively high particle mass. The OverWhelmingly 
Large Simulations (OWLS) project which consists of  a large suite of 
cosmological, smoothed particle hydrodynamics simulations with varying
 boxes and resolutions, includes AGN feedback in some of its simulations. The
highest resolution simulations ( $512^3$ particles for 
the 25 Mpc box and gas particle mass $1.4\times 10^6 h^{-1}\msun$), 
run only up to $ z=2 $ making it impossible 
to compare with the observational data on the local galaxy mass
function \citep{scha10} at this resolution. Nevertheless, the analysis of the properties of galaxies in the lower resolution OWLS
simulations ($8.7\times 10^7 h^{-1}\msun$) at  $z=0$ by \cite{muldrew13} hints at the problem of a power law stellar mass function and the absence of 
a physical scale on which the BH becomes effective. This problem is also seen in the simulations of 
\cite{puchwein13}.
Motivated by the success of semi-analytic models, \cite{sijacki2007} explicitly introduces 
 different feedback schemes for high and low mass accretion rates. This approach  along with a {\it radiative mode}, that includes  the effects of 
a strong ionising radiation emerging from active BHs in the net gas cooling rates, are used in the recent large
volume cosmological simulations of the ILLUSTRIS project \citep{vogelsberger2014, sijacki2014}.

For a cluster-scale approach, \citet{dubois2013} follow the evolution  of
a  $10^{11} \msun$ mass halo progenitor of a cluster at $z=0$, including super nova (SN) and AGN feedback. \cite{dubois2013}  found that 
the feedback from star formation plays a marginal role and that AGN feedback is able to quench star formation
when BHs are self-regulating.

In this paper we will take another look at the BH accretion models used
in these numerical simulations. In the models above, the BH
accretion rate is based on the Bondi estimate. This assumes that material falling through
the Bondi radius free-falls into the BH. As recently discussed by
several authors (e.g. \citealt{krumholz2005,power2011,debuhr2011,angles_alcazar2013}), neglecting the angular momentum of the flow may lead to
significant errors. \cite{power2011} propose a new model in which a BH particle
represents a self-regulating torus. Orbits are used to estimate whether particles are captured within a given
accretion region of the BH.  Then  gas particles are accreted onto the
BH after a viscous time. Both the accretion region and viscous time are
given by free parameters in their model. Their method is suitable for ultra high
resolution calculations (with particle mass $10^2 \msun$) where the
accretion region is $\sim 0.003 \, \kpc$, but it is not suitable for the
multi-megaparsec scale simulations that are required to study the galaxy population \citep{muldrew13}.
\cite{debuhr2011} also propose a BH accretion model that depends 
on the angular momentum. Specifically, the accretion rate is proportional to the mean gas surface density, the local sound speed squared and the 
inverse of the rotational angular frequency.  They apply their model in isolated major mergers, finding self-regulation 
of the resulting BH but no clear evidence of suppression of the star formation in the resulting galaxy. 

We  will present a model that is similar in spirit to that of \cite{debuhr2011}. We propose a simple
scheme that takes into account the angular momentum
of the gas flow, but does not require such high resolution so that it can be used in simulations of the cosmological 
population. The sub-grid model that we arrive at is suitable for inclusion in
simulations of galaxy formation, and thus allows us to revisit the interaction
of BH feedback and galaxy formation.  We show that including the
angular momentum has a profound impact on the behaviour of BH accretion,
reproducing the behaviour postulated in semi-analytic models. In contrast to
the common interpretation, however, the BH accretion shapes the
mass function not because the efficiency of BH {\it feedback} varies
with halo mass, but because BH accretion is strongly suppressed  in cold star forming disks
(relative to the rate estimated when angular momentum is not accounted for).  We show
that the new model not only matches the observed galaxy
stellar mass fractions, but also generates accretion patterns that strongly
resemble quasars in lower mass haloes and radio galaxies in the highest mass
haloes that we probe.

The strategy of the paper is as follows. Most cosmological simulations estimate BH 
accretion rates on the basis of the Bondi-Hoyle-Lyttleton accretion model \citep{bondi44}. We begin section~\ref{sec:newmodel}
by summarising this model and discuss the critical length and critical timescales in Bondi accretion. In the following 
subsection, \S\ref{sec:bhaccretionmodel}, we motivate a simple extension of the Bondi-Hoyle-Lyttleton  that allows
the model to account for the angular momentum of material flowing through the Bondi radius. This model contains 
an uncertain parameter that accounts for the effective viscosity of the disk/torus that forms as the 
flow circularises: for plausible values of this parameter, the accretion rate may be suppressed by many orders of magnitude
relative to the BS09 rate. 

In section~\ref{sec:cosmosimmodel}, we take on the challenge of the implementing the extended model as a sub-grid physics 
module suitable for cosmological scale simulations.  
To illustrate the impact of including angular momentum in the calculation,  we perform a suite of cosmological simulations. 
In section~\ref{sec:code},  we describe the simulation code, the initial conditions and the simulations
used in this study. In section~\ref{sec:galaxyproperties}, we compare the impact of BH accretion in the Bondi-like accretion model used by BS09 and the 
angular momentum dependent extension of this model. We investigate the accretion history of the most massive BHs in the simulations in section~\ref{sec:history_accretion}. Finally, we summarise our results and discuss the fundamental implications of
the BH accretion model  in section~\ref{sec:discussion}. The conclusions of the paper
are presented in section~\ref{sec:conclusions}. Convergence tests and parameter variations can be found in Appendix~\ref{appndx:convergencetests} and \ref{appndx:variation_alphavisc} respectively.

Throughout we adopt the $\Lambda$-CDM
cosmology with $h=0.704$ , $\Omega_\Lambda=0.728$, $\Omega_{\rm{m}}= 0.272$,
$\Omega_{b}=0.0455$, $\sigma_8=0.810$, and  $n_{\rm{s}}=0.967$. These values
were derived from the Wilkinson Microwave Anisotropy Probe 7 year (WMAP7) data
\citep{komatsu2011}.

\section{A Black Hole Accretion Model that Accounts for Angular Momentum}
\label{sec:newmodel} 

\begin{figure*} 
\begin{center} 
\begin{tabular}{cc}
\includegraphics[width=\columnwidth]{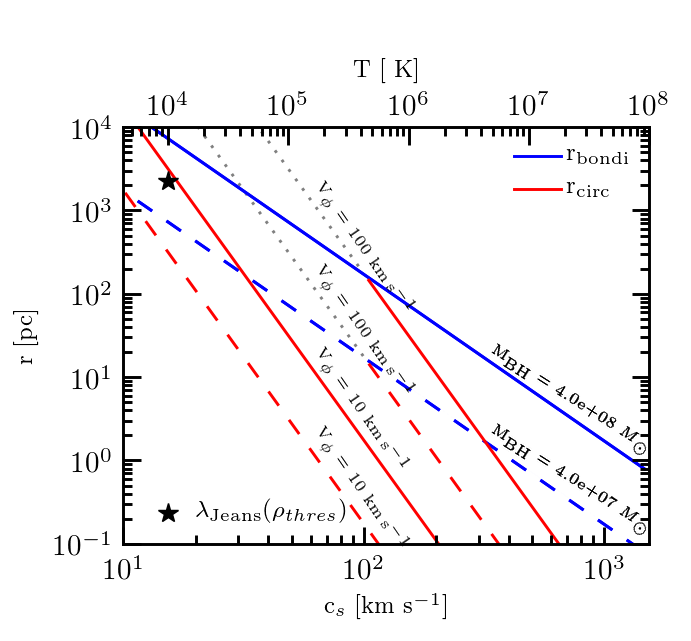} &
\includegraphics[width=\columnwidth]{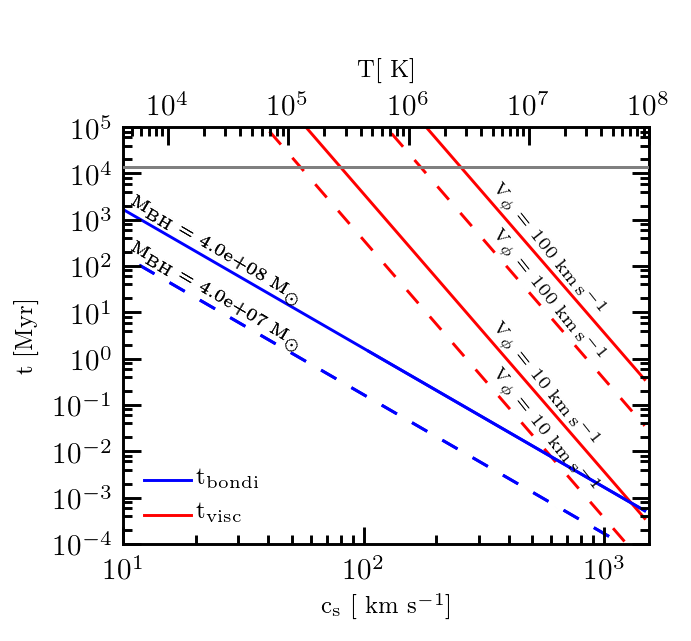} \\ 
\end{tabular} 
\end{center}
\caption{The left panel shows the Bondi (blue lines) and viscous (red lines) radii as a function
of the sound  speed for $\mbh=4\times10^7\msun$ (dashed lines) and $\mbh=4 \times 10^8\msun$ 
(solid lines) and for $\Vphi= 10 { \kms}$ and $100 {\kms}$. The grey dotted lines show the unstable regime in which
our assumption that $(\Vphi/\cs) \le 1$ breaks down. Although both radii
decrease with the sound speed and increase with $\mbh$, the circularisation radius is a
stronger function of the sound speed. The star marker represents
$\lambda_{\rm Jeans}$ in a gas at $T=10^4 \K$ and at the SF threshold density 
($n^*_{\rm H} =0.1 \,  {\rm cm}^{-3} $)
. The simulations are designed so that this is well resolved.
The right panel compares the Bondi time (blue lines) and viscous time
(red lines) for our fiducial model with 
$\Cvisc = 2 \times 10^6$ (Eq.~\ref{eq:tviscos}).  At low
values of $\cs$, the viscous time can be as long as the age of the Universe, 
but it reaches much shorter values when the sound speed is high.  The
Bondi time also decreases with high sound speed but the effect is
weaker: even for low circular speed ($10 \, \kms$), 
the accretion rate is limited by the viscous timescale 
unless the sound speed is as high as $\cs \sim 10^3 \, \kms$. } 
\label{fig:timescales} 
\end{figure*}

Black hole accretion models are almost universally based on the Bondi accretion rate or its extension, usually referred to as the 
Bondi-Hoyle-Lyttleton rate, that extends the model to the case of a BH moving relative to the background particles \citep{bondi44}. 
The net circulation of gas around the BH is not taken into account in either formulation.
In this section, we describe a simple, physically motivated extension that accounts for the net
angular momentum (AM) of the surrounding gas. 

\subsection{The physical scales of BH accretion}

In the absence of angular momentum, the accretion rate of the BH is determined by 
the gas density and effective sound speed at the Bondi radius, $\rbondi$. This radius, defined as the radius at which the BH gravity
dominates over the thermal and turbulent pressure of the surrounding gas, is a key physical length scale in the 
problem. It  is given by  
\be 
\rbondi = \frac{ G \mbh }{\cs^2} 
        \approx 430 \Big ( \frac{\mbh}{ 10^7 \msun} \Big ) \Big ( \frac{\cs}{ 10\, \kms} \Big )^{-2} \pc, 
\label{eq:bondiradius}
\ee where $\mbh$ is the mass of the BH and $\cs$ is the sound speed of the surrounding medium.
In the Bondi estimate of the accretion rate, gas within $\rbondi$ falls directly into
the BH. However, when the infalling gas has a net angular momentum, the flow
will instead form an accretion disk or torus with characteristic radius, $\rcirc$. For any significant angular momentum,
this radius will be larger than the last stable orbit of the BH. The radius of the disk is determined by the circularisation radius of the flow 
\be 
\rcirc  =  \frac{j^2(\rcirc)}{ G\mbh}     
\label{eq:rcirc1} 
\ee 
where $j$ is the specific angular momentum of material in a circular orbit at the circularisation radius. The estimate assumes that 
the mass of the BH dominates within the circularisation radius. In order
to be accreted by the BH, material within this radius must transfer its angular momemtum outwards faster than it is
converted into stars or expelled from the accretion region by feedback processes.

To produce a simple model, we assume that the accretion processes can be treated in two
parts, firstly an almost radial infall from the Bondi radius to the
circularisation radius and then a slower flow of material through the disk to
the last stable orbit.  We express the specific angular momentum of the flow passing through
$\rbondi$ in terms of the Bondi radius and the net tangential velocity (or circulation speed, $\Vphi$) of gas at the Bondi radius 
, $\bmath{j}={ \bmath \rbondi}\times {\bmath \Vphi}$. Assuming angular momentum is conserved in the infall
phase, the circularisation radius is
\bary 
\rcirc &=&\frac{ \rbondi^2 \Vphi(\rbondi)^2 }{G\mbh}  = G\mbh\frac{ \Vphi(\rbondi)^2}{\cs^4} \nonumber  \\
       &\approx&430\Big (\frac{\mbh}{10^7 \msun} \Big ) \Big ( \frac{\Vphi(\rbondi)}{10 \, \kms}\Big )^2 \Big( \frac{\cs}{10\,  \kms } \Big)^{-4} \pc .
        \nonumber \\ 
\label{eq:rcirc2}
\eary 
This formulation implies that $(\rcirc/\rbondi) \le 1$, since material at the Bondi radius is assumed to be falling into the BH. This implies $(\Vphi/\cs) \le 1$ {\em at the Bondi radius}.  On scales larger than the Bondi radius, it is entirely possible that the circulation speed will be greater than the sound speed.  This will be the case, for example, in a rotationally supported cold gas disk.

To get a sense of the values of the Bondi and circularisation radii relative to the Jeans scale of the interstellar medium we include Fig.~\ref{fig:timescales}.  
The left panel illustrates the variation of $\rcirc$ (red lines) and $\rbondi$ (blue lines) with the circular speed and
the effective sound speed  (or equivalently, the {\it effective temperature}) of the surrounding gas\footnote{
We define the effective sound speed as $\cs \equiv \Big ( \frac{\gamma k_B T}{\mu m_H}\frac{1}{f_{ \rm th}} \Big)^{1/2}$  
where $ f_{\rm th}$ is the fraction of pressure which is thermal. In our simulations, we assume $f_{\rm th}=1$ but adopt an effective gas temperature that is given by 
the ISM equation of state (see BS09). We set $\mu=0.59$.}. We limit coloured lines to the region $(\Vphi/\cs) \le 1$, where the circularisation radius is smaller than the Bondi radius 
as required for infalling material.

The left panel in Fig.~\ref{fig:timescales} shows $\rbondi$ and $\rcirc$ for  two values of the BH mass $\mbh$ 
at $4\times 10^8$ (solid lines) and $4 \times 10^7 \msun$ (dotted lines) and two values for the circular speed $\Vphi$,
$10$ and $100$ $\kms$. Both  the Bondi and circularisation radii
vary  from thousands of parsecs when the sound speed  is $10 \, \kms$ (gas
temperature $ \sim 4 \times 10^3 \, \K$)  to less than 1 parsec when the sound speed is
$10^3$ $\kms$ (gas temperature $\sim 4 \times 10^7 \, \K $). Although both radii decline as the sound speed increases, the
circularisation radius shows a much more dramatic variation so that the importance of the accretion disk phase is reduced relative to the 
free-fall phase of the flow.
  	
It is useful to compare the Jeans length $\lambda_{\rm Jeans}$ ($\sim 2.2 \,  \kpc$, black star)  for the gas  at the star formation density 
threshold ($n_{\rm H}^* = 0.1\,  {\rm cm}^{-3}$) with an effective temperature $10^4 \, \K$  \footnote{ There is no unique 
definition of the Jeans length. This usually differs by a factor 
which depends on the geometry of the object. We define   
$\lambda_{\rm Jeans}\equiv \Big(\frac{\cs^2} { G \rho} \Big)^{1/2}=\cs { n_{\rm H}^{*}}^{-1/2} 
\Big(\frac{ X f_{gas}}{  G \mu m_{H}}\Big)^{1/2} $ where $f_{gas}=0.3$ is the gas mass fraction and $ X=0.752$ is 
the hydrogen mass fraction. Setting $\cs \sim 15.3 \, \kms $,  $\lambda_{ \rm Jeans} \sim 2.2$ kpc. 
We define the corresponding Jeans mass as $ M_{\rm Jeans}=4/3 \pi \rho \left(\frac{\lambda_{\rm Jeans}}{2}\right)^3 $  that takes
the value $2 \times 10^7 \msun$.}  For a BH with mass comparable to the Jeans mass ($2\times 10^7 \msun$), the Bondi radius 
is just less than a kpc, similar to the Jeans length.  In practice, however, gas surrounding the BH will usually be far denser
than the star formation threshold. Turbulence, induced by star formation (through the winds of hot stars and supernovae),
will greatly increase the effective sound speed and consequently reduce the Bondi radius. Assuming the net circulation speed is independent of radius, 
the smaller Bondi radius implies a lower angular momentum for the inflowing gas and consequently a smaller circularisation radius.


\subsubsection{The Bondi timescale and the viscous timescale}

In the previous section, we showed that the presence of angular momentum introduces 
an additional spatial scale into the accretion model, corresponding to the radius at which infalling material circularises.
This creates an additional timescale that is dependent on the angular momentum of the infalling material. 

In the case that the gas around the BH  does not have any net circulation, 
the timescale for gas accretion is determined by the sound-crossing time at the Bondi radius:
\bary 
\tbondi & = &\frac{\rbondi}{\cs} = \frac{G\mbh}{\cs^3} \nonumber \\
        &\approx& 42 \Big (\frac{ \mbh}{10^7 \msun} \Big ) \Big (\frac{\cs}{ 10\,\kms} \Big )^{-3}  \Myr. 
\label{eq:tbondi} 
\eary
In the presence of a net circulation, an accretion disk or torus will be formed at the circularisation radius. The disk viscosity
will cause gas to spiral inwards to the last
stable orbit where it will finally be accreted. To estimate this 
timescale, we will assume that the disk formed is thin (i.e. that the 
gas cools efficiently)  and that the radial motions are small in comparison to the rotation. In this case,
gas follows roughly Keplerian orbits and the disk scale height $ H$ is given by
\be 
\frac{H}{R} \sim \frac{\cs'(\rcirc) }{ v'_{\rm circ}(\rcirc)} = \frac{\Vphi}{\cs}\frac{\cs'}{\cs} \ll 1, 
\label{eq:scaleheight} 
\ee 
where $ v'_{\rm circ}$ is the circular velocity within the disk, and
$\cs'$ is its sound speed. The second equality follows from  conservation  of specific angular momentum between the circularisation radius
and the Bondi radius (ie., $\rcirc v'_{\rm circ}(\rcirc) = \rbondi \Vphi(\rbondi)$) using Eqs.~\ref{eq:bondiradius} and~\ref{eq:rcirc2}. 
To be consistent with these assumptions and that of 
a thin disk, $\cs' \ll \cs^2/\Vphi$, which follows if there is strong turbulence in the ISM.
Transport through the disk can be  described
by a diffusion equation where the kinematic viscosity $\nu$ is usually
parameterized as 
\be 
\nu =\alpha_{\rm visc} \, \cs' H,
\label{eq:kineticviscosity} 
\ee 
with $\alpha_{\rm visc}$ as a dimensionless number \citep{shak73}.  The viscous time can then be expressed as 
\be 
\tvisc = [\alpha_{\rm visc} (H/R)^2]^{-1}  t_{\rm dyn}\sim \frac{\rcirc^2}{\nu},
\label{eq:tvisc0} 
\ee 
where $ t_{\rm dyn}$ is the dynamical timescale of the disk. The values for $ H/R$ lie 
in the range $0.1-0.001$ for a thin accretion disk in the $\alpha$-disk parameterization \citep{shak73} 
and values of  $\alpha_{ \rm visc}$ lie in the range $\sim 0.1-0.3$ from observational evidence
\citep{king2007,schreiber2004,buat2001,canizzoa01,canizzob01}.  Using
Eqs.  ~\ref{eq:rcirc1}, ~\ref{eq:rcirc2}, ~\ref{eq:kineticviscosity} and ~\ref{eq:tvisc0}, the viscous time becomes 
\bary 
\tvisc &=& \Cvisc \frac{\rcirc}{ v_{\rm circ}} = \Cvisc \frac{j^3}{ G^2\mbh^2}  \nonumber \\
       &=& \Cvisc  \frac{\rbondi^3 \Vphi^3}{  G^2\mbh^2} =\Cvisc  G\mbh\frac{\Vphi ^3}{\cs ^6} \nonumber \\
 &\approx& \resizebox{.75\hsize}{!}{$42\,\Cvisc\big(\frac{\mbh}{10^7 \msun}\big) \big ( \frac{\Vphi}{ 10\,\kms} \big)^{3} \big( \frac{\cs}{10 \kms} \big)^{-6} \Myr,$} 
\label{eq:tviscos} 
\eary 
For a Shakura-Sunyaev disk, $\Cvisc= 2\pi[\alpha_{visc}(H/R)^2 ] ^{-1}$.  If we fully understood the transport processes in the rotating disk or torus that feeds the BH, we would simply insert the appropriate values for  $\alpha_{visc}$ and $(H/R)$. For a thin disk model, plausible values of $\Cvisc$ lie in the range $10^8  - 10^3$.
However, the structure of the circularisation disk we are considering is completely unclear, as is the dominant viscous mechanism on 
relevant scales. These issues are discussed extensively in the literature (e.g. \citealt{shlosman1990,king2011, power2011, hopkins_quataert2010}) because the long timescales implied 
by the Shakura-Sunyaev formulation make it hard to understand the high efficiency of accretion required to create supermassive 
BHs at very high redshift. On the scales relevant to our simulations, the appropriate transport mechanism is likely to be 
gravitational instability \citep{hopkins_quataert2010, hopkins_quataert2011} rather than the magneto-rotational instability (MRI). 
But these simulations do not accurately treat the multiphase structure 
of the ISM and stellar feedback, and
the value of $\Cvisc$ is therefore extremely uncertain. We adopt an empirical approach by varying $\Cvisc$ to obtain the best match to 
observed galaxy properties, as discussed in appendix~\ref{appndx:variation_alphavisc}.  The analysis prompts us to choose $\Cvisc =2 \times 10^6$ as a fiducial value, 
although a wide range of values give qualitatively similar results.

The right panel of Fig.~\ref{fig:timescales} shows $\tbondi$ (blue lines)
and $\tvisc$ (red lines) as a function of the effective sound speed (or
equivalent gas temperature) for BH  masses $\mbh= 4\times 10^{7}\, \msun$ and
$\mbh= 4\times10^8 \,\msun$ and $\Cvisc =2.1 \times 10^6$ under the same region of our assumption ( $\Vphi/\cs\le 1$).  The Bondi time and viscous time are both decreasing
functions of the sound speed, but the viscous time has a stronger dependence on the sound speed. For instance, for a BH with  
$\mbh= 4\times 10^8 \, \msun$ surrounded by gas with $ \cs\sim 15.3 \, \kms$, the Bondi timescale is $\sim 1\,  \Gyr $,
but this drops by six orders of magnitude (to $\sim 10^3 \,  \yr $) 
when $\cs \sim 10^3 \,  \kms$. It is important to note, however, that although the accretion timescale is much shorter when $c_s$ is large, the overall accretion
rate depends on the local density: at fixed pressure, the accretion rate declines with increasing $c_s$.
For $\tvisc$, the sound speed dependence is even stronger: For a circular speed of $10\,  \kms$, the viscous time is $ \sim 10\, \Gyr$ 
($\sim$ the age of the Universe) for $\cs \sim 90 \,  \kms$,  and  $\sim  10^3 \,  \yr$  at $\cs \sim 10^4 \, \kms$ . In contrast, both timescales are 
relatively weak functions of the BH mass. The figure  particularly highlights that the mass accretion rate will be 
mostly predominantly limited by the viscous time because of the stronger dependence of $\tvisc$ on the local sound speed.


\subsubsection{BH accretion rate accounting for angular momentum}
\label{sec:bhaccretionmodel}

Material with sufficiently low specific angular momentum falls through the Bondi radius forming an angular momentum supported disk 
or torus. This leads to a dependency of  the viscous time and circularisation radius  on the effective sound speed  and the circular velocity 
of gas flowing through the Bondi radius. In this section we propose a
revised estimate of the mass accretion rates of BHs that takes gas circulation into account.

The long viscous timescale of the accretion disk creates a bottleneck for accretion onto the BH. As matter flows through the Bondi radius it will pile up at the circulation radius. If the viscous timescale is long, this matter will form a nuclear star burst with only a small fraction being accreted into the BH. The challenge is to estimate the mass of the accretion torus/disk, and thus the accretion rate. Our subgrid implementation requires an estimate of the instantaneous accretion rate: gas that has stalled will be represented by macroscopic simulation particles (star formation and winds are handled by other components of the simulation code).  The Bondi time sets the characteristic timescale of the problem. In the absence of circulation, the mass within the Bondi region is $\sim \mdotbondi  \tbondi$.
We assume that the viscous bottleneck results in this mass building up in the disk/torus and then draining into the BH at a lower rate:
\be
\mdotbh \sim  \frac{\mdotbondi  \tbondi}{ \tvisc}.
\ee
The constant of proportionality is degenerate with the proportionality constant $\Cvisc$. 
Following this argument, the critical factor is, then, the ratio of the Bondi and viscous times:
\bary
\frac{\tbondi}{\tvisc} & = & \frac{\rbondi \cs^{-1}}{\Cvisc [ \rbondi \Vphi]^3 [ G\mbh ]^{-2}} \nonumber\\
                       & = & \frac{1}{\Cvisc}\frac{\cs^3}{\Vphi^3}. 
\label{eq:suppressionfactor}
\eary 
Thus $\tbondi/\tvisc$ depends only on $[\cs / \Vphi]^3$.  Note that
if $\tvisc$ is larger  than $\tbondi$ (i.e.  $\Cvisc^{1/3}\Vphi>\cs$) the accretion rate is 
limited by the Bondi rate and we can ignore the time spent in the accretion disk phase. We therefore
write the BH  accretion rate as  
\begin{equation}
\mdotbh = \left \{ 
             \begin{array}{cc}
              \dot{m}_{\rm Bondi} \,\left[ \frac{1}{\Cvisc}\big(\frac{\cs}{\Vphi}\big)^3 \right] 
                                                   &  {\rm if~} \Cvisc^{1/3}\Vphi> \cs, \\ 
              \dot{m}_{\rm Bondi} &  \textrm{Otherwise} \\ 
             \end{array} 
          \right. 
\label{eq:newbhacr}
\end{equation}
where $\dot{m}_{Bondi}$ is the mass accretion rate given in the Eq.~
\ref{eq:bh1} corresponding to the Bondi-Hoyle-Lyttleton accretion model
(following \citealt{bondi44} to allow for the bulk gas motion as well as gas
circulation) and $\Cvisc$ is a free parameter which parameterizes the
efficiency of angular momentum transport and mass loss from the disk. We take
$\Cvisc=2.1\times10^6$ as our fiducial value. 


The circular speed of material at the Bondi radius needs to exceed a critical
value before the accretion timescale is suppressed. In the case of our
fiducial, $\Cvisc=2.1\times 10^6$,  this critical value is given by $\Vphi
\approx\cs /128$.  For sound speeds of $10^3 \, \kms$, the critical value of
$\Vphi$ is $10 \, \kms$ and for low sound speeds of $10 \,  \kms$  it is  $0.1
\,  \kms$.  Below  this critical value, even though a disk/torus forms,
transport through the disk is more rapid than the rate at which material flows
through the Bondi radius.

\section{BH accretion modelling in cosmological simulations}

\label{sec:cosmosimmodel} 
Most cosmological simulations include accretion onto central BHs via a sub-grid
model based on the Bondi-Hoyle-Lyttleton accretion model \citep{bondi44},
possibly adding a coefficient that attempts to account for the multiphase
nature of gas around the BH that is not resolved at the finite
resolution of the simulation (e.g. \citealt{springel2005};
\citealt{dimatteo2005}; BS09). These models assume that the net circulation of
gas in the neighbourhood of the BH can be neglected.  In this section we will
discuss how these models can be extended to account for the circulation of gas.
We focus our description on the model of BS09, but the extension can equally be
applied to other implementations of the Bondi-Hoyle-Lyttleton accretion model.
It is inherent in the nature of sub-models that the density, effective sound
speed and circulation of gas at the Bondi radius will need to be estimated from
coarse scale quantities that are resolved in the simulation.

\subsection{The OWLS BH accretion model} 
\label{sec:owlsmodel}

Following \cite{springel2005}, a BH seed with a mass $\mseed$  is
injected into each new Friends-Of-Friends halo that exceeds the mass threshold,
$m_{halo,min}$. In order to prevent the BH from being ejected by two-body
encounters, the BH seed is assumed  to track the position of the minimum
potential in the halo.  Subsequently, BHs grow by two processes: mergers with
other BHs or accretion of nearby gas particles. The gas accretion obeys  a
modified version of the Bondi-Hoyle-Lyttleton formula: 
\begin{equation}
 \dot{m}_{\rm Bondi} = \alpha \frac{4\pi G^2 \mbh^2\rho}{(\cs^2+v^2)^{3/2}}, 
\label{eq:bh1} 
\end{equation} 
where $\mbh$ is the mass of the BH, $\cs$ is the sound speed, $\rho$
density of the local medium, $v$ is the velocity of the BH relative to the
ambient medium. The coefficient $\alpha$ is a dimensionless efficiency parameter that is
introduced to account for the multiphase nature of the gas around the BH. 
 The complex density structure of the ISM cannot be explicitly realised unless the simulations have a resolution significantly 
better than a few pc \citep{creasey12}.  In the original \cite{springel2005} model, $\alpha$ is a fixed
number that ensures that the BHs grow rapidly during galaxy mergers. The authors set $\alpha=100$. 
In BS09, the structure of the dense star forming ISM is treated using an imposed
polytropic equation of state that matched on to the ideal gas law at a threshold density, $n_{\rm H}^*$ \citep{scha08}.
Gas at lower densities is assumed to be single phase, and the macroscopic density to therefore be a good estimate of the 
gas density at the Bondi radius of the BH.  BS09 therefore extend the model, 
so that $\alpha$ becomes a function of local density,  
\begin{equation} 
\alpha=\left \{ \begin{array}{cc}  1 &  {\rm if~} n_{\rm H}<n_{\rm H}^{*} \\ 
                         \Big(\frac{n_{\rm H}}{n_{\rm H}^*}\Big)^\beta
                                    &\textrm{otherwise,} 
\end{array}
\right. 
\label{eq:beta} 
\end{equation} 
BS09 show that the parameter $\beta$ does not play a crucial
role in the  growth of BHs which is usually limited by the AGN feedback. We set the parameter $\beta$ to 2 which 
is the fiducial value in the OWLS project.  We will refer to Eqs.~\ref{eq:bh1} and \ref{eq:beta} as the Bondi-like 
model in what follows.

The sub-grid accretion model also assumes that the BH accretion rate cannot exceed the Eddington accretion rate limit,
\begin{equation}
 \dot{M}_{\rm Edd} = \frac{4 \pi  G m_{\rm p} \mbh}{ \sigma_{\rm T} \, c \, \epsr},
\label{eq:accr_edd}
\end{equation}
where $ m_{\rm p}$ is the proton mass, $\sigma_{\rm T}$ is the Thomson cross section for the scattering 
of free electrons  and $\epsf$ is the fraction of the energy liberated per accreted rest mass 
energy and is set at $0.1$. The final accretion rate is 
\begin{equation}
{ \dot{m}}_{\rm BH} = {\rm min}(  \dot{m}_{\rm Bondi} , { \dot{M}}_{\rm Edd}). 
\end{equation}

\subsection{Accounting for angular momentum (AM) dependent accretion
  on the sub-grid model}
\label{sec:implementation} 

In order to implement the AM dependent accretion rate as
a sub-grid model, we need to estimate the circulation speed of the gas surrounding a BH using the SPH smoothing kernel.  We
determine the weighted AM of the surrounding gas and then divide by the smoothing length $h$: 
\be 
\Vphi = \Big | \sum_{i=0} ^{ N_{ \rm SPH}} \mathbf{r_i}\times \mathbf{ v_i} \, m_i W(\mathbf{r_i},{h})\frac{1}{\rho h} \Big| 
\label{eq:circspeed} 
\ee 
where $ W(r,h)$ is the SPH smoothing kernel used in a SPH-code, $h$ is the SPH smoothing
length of the BH and $\rho$ is the smoothed density given by $\rho =
\sum_{i=0}^{N_{\rm SPH}} m_{i} W(r_i,h)$. 

The cosmological simulations  almost never resolve the Bondi radius, so the sub-grid scheme must 
extrapolate the circularisation speed that is measured to smaller scales, just as it is needed to extrapolate the resolved gas density 
in the neighbourhood of the BH to the density at the Bondi radius. Following a similar approach, $\Vphi(\rbondi)$  should 
be extrapolated from the resolved circulation as a follows 
\be
\Vphi( \rbondi)=(\frac{\rbondi}{h})^{\gamma} \Vphi( h ),
\ee
where $\gamma$ can, in principle, be positive or negative depending on the relative mass of the BH and the structure of the surrounding gas. 
For simplicity, we will set $\gamma=0$  in this paper.  This assumes a flat rotation curve in the region around the BH and those 
processes that disrupt the rotation of the accretion disk on resolved scales would also disrupt the circulation on the scale of 
the Bondi radius.  Note that the choice of  $\gamma$ is largely degenerate with $\Cvisc$,  so that if a different value of $\gamma$ was
chosen, $\Cvisc$ would have to be re-calibrated. 

In a cold, rotationally supported gas disk, $(\Vphi/\cs) \ge 1$ on the smallest scales
that we are able to resolve.  In this situation, the circularisation lies outside the Bondi radius and we continue to apply Eq.~\ref{eq:tviscos} as the viscous timescale. Clearly in future work it would be possible to develop a more complex model in which the viscous timescale had a more complex dependence on the spatial scale
of the circularisation radius. Our present approach is intended to be as simple as possible.

The subgrid model provides an estimate of the instantaneous accretion rate. Following \cite{springel2005}, BH particles are described by two masses, a subgrid mass used in the accretion rate calculation and a particle mass 
used in gravitational 
calculations. This allows the subgrid BH to be less massive than the particle mass used in the simulation,
but ensures that gravitating mass is strictly conserved. If the BH subgrid mass exceeds the particle mass, 
a particle within the BH neighbourhood is stochastically selected and its mass added to the particle
mass of the BH particle. This allows the two BH masses to track each other. The surrounding 
gas particles continue to form star and potentially generate outflows.

\begin{figure} 
\begin{center} 
\includegraphics[width=7.5cm]{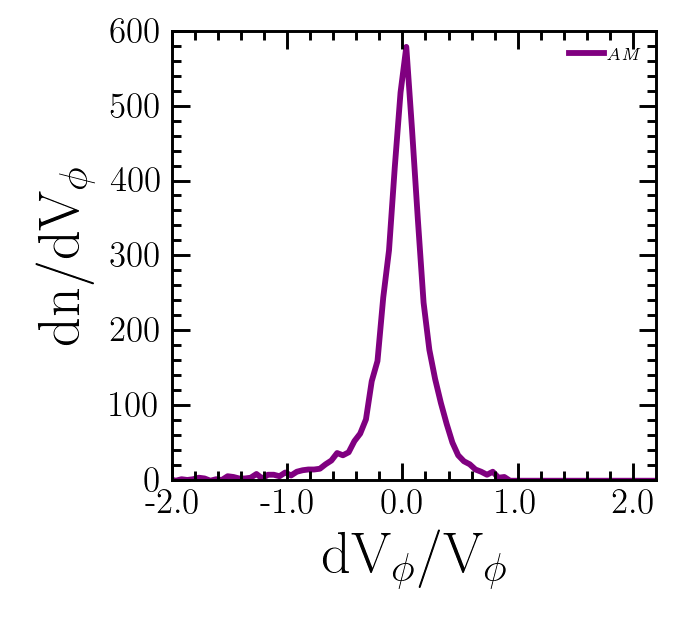} \\
\end{center}
\caption{ The normalized distribution of fractional fluctuations of $\Vphi$ , ${\rm d} \Vphi/\Vphi $ for the BHs in the simulation {\it AM} (see section \ref{sec:code} ) . In order to place an upper-limit on the 
numerical noise in the measurement of $\Vphi$, we compute the fractional change between consecutive time steps. This variation is an upper
limit on numerical nose since the fluctuations may be dominated physical variations in the gas flows near to the BH (driven, for example, by BH  and stellar feedback). The form of the histogram is described well by a Cauchy
distribution with a Gaussian equivalent $\sigma$ of $0.20$\label{fig:deltavphi_distribution}} 
\end{figure}

Calculating the circular velocity using Eq.~\ref{eq:circspeed} gives stable values of $\Vphi$ even though the
smoothing length and AM may fluctuate wildly. 
In order to set an upper limit on the numerical fluctuations in $\Vphi$, we calculate the relative change in $\Vphi$,  ${\rm d} \Vphi/\Vphi$,  between 
consecutive timesteps. Fig.~\ref{fig:deltavphi_distribution} shows the combined distribution for each BH with final mass $>10^{6}\msun$. 
The distribution of $\rm d \Vphi/\Vphi$ is well characterised by a Cauchy distribution, we use half the 16--84 percentile range to determine
a Gaussian-equivalent $\sigma$ of $0.20$. The distribution includes physically driven changes in the accretion rate, but the width of 
the core allows us to set an upper limit on the numerical noise arising from the SPH averaging of  less than $\sim 20\%$.  This variation is likely
to be dominated by physical gas flows (as opposed to numerical noise) driven, for example, by secular evolution and feedback from stars
and the BH.  We also find that the estimated value of the circular speed is largely independent of the
number of neighbours used in the smoothing kernel.  Furthermore, in Appendix~\ref{appndx:idealised_galaxies}, we show that the
values of $\Vphi$ and $\cs$ converge well with increasing particle number in idealised galaxy simulations.  Note that our approach is very different from
\cite{debuhr2011}, who estimate the circular speed using the gravitational mass only. This difference is 
important: variations in the circular speed (driven by local gas conditions) result
in distinctive accretion patterns for stable disks and merger induced outbursts.  The circular speed
is then used to compute the viscous timescale using Eq.~\ref{eq:tviscos}, and
thus the modified accretion rate using Eq.~\ref{eq:newbhacr}. The accretion
rate is also constrained to be smaller than the Eddington accretion rate (see Eq.~\ref{eq:accr_edd}).  
Other aspects of the implementation are identical to \cite{booth_schaye09} which we will describe 
briefly in section~\ref{sec:code}.

\section{The Numerical Code and Hydrodynamic Simulations}

\label{sec:code}

The simulations we present are based on the GADGET-3 SPH code
\citep{springel2005b}, adding enhancements to reduce the simulation viscosity
when the time derivative of the flow divergence is small
\citep{cullen_dehnen2010}
 and to ensure that timesteps of particles receiving feedback energy
are limited \citep{durier_dallavecchia12}. These enhancements will be
described in more detail in Dalla Vecchia (in prep). The code uses an extensive
network of sub-grid modules to account for the turbulent pressure of the ISM,
and implements star formation following the empirical Kennicutt-Schmidt law
\citep{scha08}, chemical enrichment \citep{wiersma2009b} and cooling tracking 11
elements \citep{wiersma2009a}. These modules were originally
developed for the OWLS \citep{scha10} and GIMIC \citep{crain09} simulation
projects. Here we use the metallicity dependent gas density threshold of \cite{scha04}
(as in OWLS model SFTHRESZ) and a revised treatment of the equation of state \citep{dallavecchia_schaye12}.
Feedback from stars is implemented as stochastic thermal energy injection
events, using a fixed heating temperature of $10^{7.5}\, \K$, in order to avoid
spurious radiative losses (\citealt{dallavecchia_schaye12}, see also \citealt{creasey11}).
We moderate the supernova feedback 
efficiency, $f_E$ (the fraction of SN energy available to perform feedback) as a function of the local dark matter velocity dispersion
\citep{oppenheimer06,okamoto08} in order to obtain a good match to the 
 faint end slope of the galaxy stellar mass function.  A similar variation is commonly used in
semi-analytic models (e.g. \citealt{guo2011,bower2012}) and is supported by
small-scale simulations of galaxy winds such as \cite{creasey12}. The feedback
efficiency $ f_E$
is set as a function of the equivalent halo virial temperature,
varying between 1.0 (in low temperature haloes) and 0.1 (in high 
temperature systems), following the equation:
 \begin{equation}
 f_E =  1.0 - 0.9\left(1 \over 1 + \exp\left[-2(\log T-\log T_0)\right]\right).
\end{equation}
After some experimentation, we set the transition temperature to $T_0 =10^{5.5} \K$.
 
In this paper, our focus is the interaction between feedback from BH
accretion and the formation of galaxies. We fix the supernova feedback
efficiency and its scaling and vary the parameters of BH feedback.
Throughout the paper, for AGN feedback  we use the stochastic heating model of BS09 with a
heating temperature of $10^8 \K$, and a density power $\beta=2$ in
Eq.~\ref{eq:beta}. By default we assume that the energy liberated per accreted 
rest mass energy is $\epsr =0.1$, and 
that the heating efficiency (this is  the fraction of liberated energy coupled to the surrounding gas medium) is $\epsf=0.15$; we 
show the effect of reducing the heating efficiency in sec.~\ref{sec:galaxyproperties}.  Accretion is always limited
to be less than the Eddington accretion rate.  BH seeds are inserted
into haloes that exceed a Friends-Of-Friends halo mass of $10^{10}  h^{-1}\msun$,
corresponding to $\sim 1500$ dark matter particles. Such haloes are well defined and
there is no danger of injecting BHs into spurious systems
(BS09). BH are injected with an initial mass of $10^4 {h^{-1}}\msun$. 
They are allowed to merge within their SPH smoothing kernel   and have relative velocity less than $0.5$ of 
the sound speed of the surrounding gas. In section~\ref{sec:implementation}, we describe the enhancement of the standard accretion 
model to account for the circular speed of gas within the smoothing kernel of the BH.

\begin{table}
\caption{A list of the simulations used in this paper. Each simulation
  has the same supernova feedback parameters, and a co-moving volume of $(25 \,\, \Mpc)^3$. 
The simulation use $2 \times 360^3$ particles, with initial baryonic particle mass
$1.4 \times 10^6  { h^{-1}}\msun$ and dark matter mass particle 
$6.3 \times 10^6 { h^{-1}} \msun$; the mass of seed BHs is set  $\mseed =10^4  h^{-1}\msun $ and 
the minimum halo mass in which BH seeds are injected is $10^{10} h^{-1}
\msun$. The columns show: 
1) Name of the simulation; 2) Efficiency with which
  energy emitted by a BH is coupled into the ambient gas; 3) Radiative
  efficiency of BH accretion disk }
\begin{tabular}{|l|l|l|p{3cm}|}
\hline
Name& $\epsf$ & $\epsr$  & Accretion model \\
(1) & (2) & (3) & (4)  \\
\hline
{\it NO-AGN}    &   --   & --  &   --                    
\\
{\it BS09}     &    0.15& 0.1 &  Bondi-like accretion model  Eqs.~\ref{eq:bh1} and \ref{eq:beta}.  
\\
{\it BS09-LE} &  0.015 & 0.1 &  Bondi-like accretion model Eqs.~\ref{eq:bh1} and \ref{eq:beta}. 
\\
{\it AM }        &   0.15 &  0.1 & AM accretion model  Eq.\ref{eq:newbhacr}, $\Cvisc =2\times10^6$. 
\\
{\it Appendix \ref{appndx:variation_alphavisc} }&  0.15 &  0.1 & AM accretion model $\Cvisc =6\times10^4$ -- $6\times10^6$ . 
\\
\hline
\end{tabular}
\label{table:simulations}
\end{table}

The simulations presented in this paper are summarised in 
Table~\ref{table:simulations}. All the simulations use the same
sub-grid parameters to describe star formation and stellar feedback,
and only differ in the BH sub-grid physics. There are 
in total 4 simulations : \begin{itemize}  
 \item {\it NO-AGN}: This simulation does not include BH physics.
 
 \item {\it BS09 }: We use BS09's model with the
   feedback efficiency set to $\epsf=0.15$ in the Bondi-like accretion
   Eqs.~\ref{eq:bh1} and \ref{eq:beta}. Accretion is also limited by the Eddington rate.

 \item {\it BS09-LE }: We set this simulation as the above except  that the feedback 
          efficiency is set to $\epsf=0.015$.
          
 \item {\it AM}: This simulation uses the AM dependent accretion model Eq.\ref{eq:newbhacr}, with $\Cvisc = 2.1 \times 10^6$.
\end{itemize}
The results are not particularly sensitive to the moderate choice of $\Cvisc$ (by 2 orders of magnitude), and we show some example variations in Appendix~\ref{appndx:variation_alphavisc}.
All of the models assume that the accretion rate cannot exceed the Eddington limit (Eq.~\ref{eq:accr_edd}).

Initial conditions were generated using second-order Lagrangian perturbations \citep{jenkins2010} at a starting
redshift of $127$.  We use initial gas and dark matter particle masses of $1.4 \times 10^6  { h^{-1}}\msun$ 
and $6.3 \times 10^6 { h^{-1}} \msun$ respectively; the {\it co-moving} gravitational softening lengths are 
set to $2  \, h^{-1} \,\kpc$ with a maximum physical softening of $  0.5\,h^{-1}\, \kpc$. The simulations are carried out in 
periodic boxes of $25 \, \Mpc$ on each side. The largest structure that forms
in the simulation has a mass of $1.65 \times 10^{13} \msun$ at $z=0$. Comparing to large simulation 
volumes we find that a box of this size does not noticeably bias galaxy properties as a function of 
virialised halo mass \citep{haas2012}, allowing us to efficiently test the impact of different 
sub-grid models. Because of the limited box size, we compare galaxy properties to the stellar fractions ($f_{\rm star} = M_{*}/\mcrit$)
inferred from observations as a function of halo mass, rather than to the galaxy stellar mass function itself.

\section{The global impact of BH accretion on galaxies}
\label{sec:galaxyproperties}
In this section we examine the impact of BH feedback on the galaxies 
in the simulations described above.  We focus on the stellar mass fraction--$\mcrit$ (halo mass) relation and  
the $\mbh$--$\mcrit$ relation, in particular.  We first present results based on the 
Bondi--like models ({\it BS09} and {\it BS09-LE}), and then contrast them with the AM dependent model.

\begin{figure*} 
\begin{tabular}{cc}
\includegraphics[width=\columnwidth]{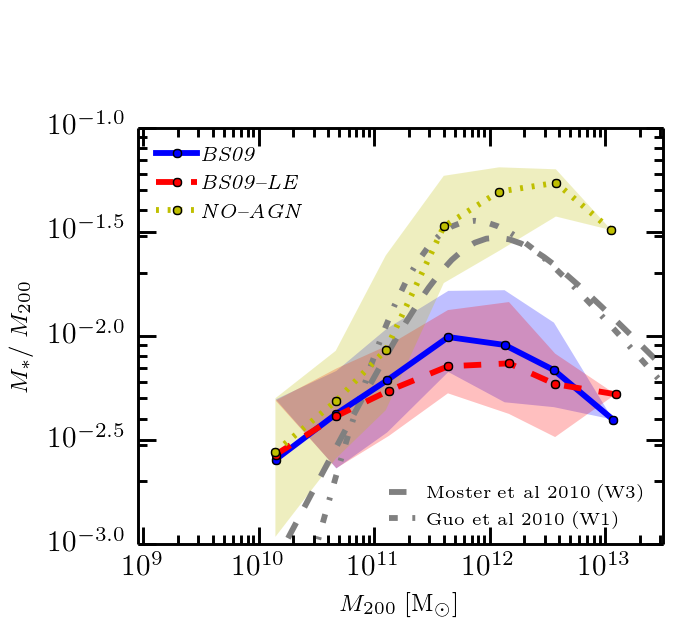} &
\includegraphics[width=\columnwidth]{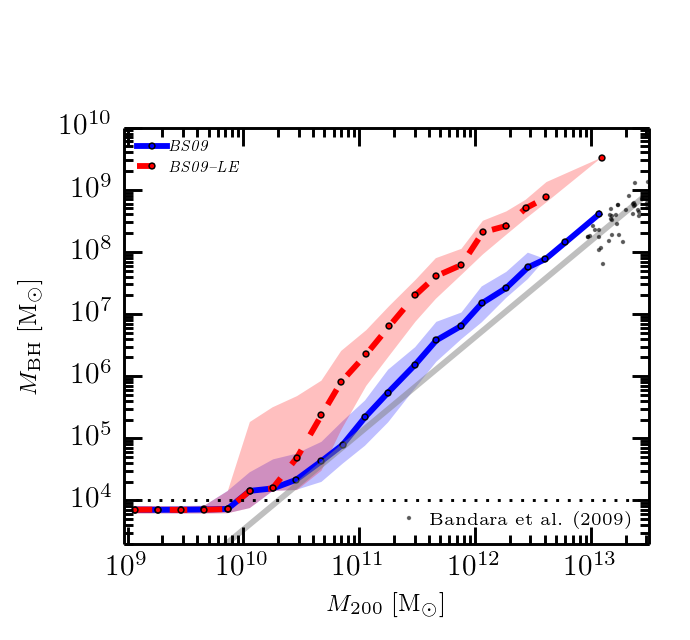} \\
\end{tabular} 
\caption{ The left-hand panel shows the stellar mass fraction as a
  function of $\mcrit$, while
  the right panel shows the $\mbh$--$\mcrit$ relation.
The lines show the medians of $M_{*}/\mcrit$ (or $\mbh$) for the
simulations {\it NO-AGN} (yellow, dotted line), {\it BS09} (blue, solid line ) and {\it BS09-LE} (red, dashed line) (see
table~\ref{table:simulations}). The shaded region shows the 10-90 percentile
range of the data. The grey dashed and
dashed--dotted lines in the left panel  represent abundance matching results derived from 
\citet{moster10} and \citet{guo10} respectively and the grey line  in the right panel indicates 
the observational $\mbh$--$\mcrit$  relation derived from observations by \citet{bandara2009}. Their observational data is shows as grey points. The figure shows that
efficient BH self-regulation creates a strong correlation between BH
mass and halo mass, but strongly suppresses the formation of stars
across a wide range of halo mass. Although changing the efficiency of
BH feedback alters the normalisation of the $\mbh$-$\mcrit$
relation, it has little effect on the host galaxy properties.}
\label{fig:effeed} 
\end{figure*}

\subsection{Simulations with BS09 Bondi--like accretion models}
\label{subsec:bondimodels}
Firstly, we look at the simulations based on the Bondi-like accretion model: {\it BS09}
and {\it BS09-LE} and compare these with the {\it NO-AGN} simulation.
The left-hand panel of 
Fig.~\ref{fig:effeed}  shows the stellar mass fraction, $\mstar/\mcrit$ (where the stellar mass is measured within a $30\, \kpc$  
radius) for central galaxies
as a function of $\mcrit$. Given the small box size, this provides a convenient way to compare with observational data.
The solid lines represent the median relation in bins of halo mass, with the 10--90 percentile range 
of the data indicated as coloured regions. Grey lines
show observational estimates based on  abundance matching from \cite{moster10} and \cite{guo10}.

The {\it NO-AGN} simulation  (yellow colour) is  consistent with the abundance matching results up to a 
halo mass of $10^{12} \msun$, suggesting that in the physical Universe the feedback from AGNs has little or no effect in low-mass haloes. The lack of AGN feedback leads to an overproduction of stellar mass in above $10^{11.5} \msun$ haloes
in disagreement with the abundance matching results. Fig.~\ref{fig:effeed} also 
shows the {\it BS09} (blue line) and
{\it BS09-LE} (red line) simulations. In both simulations, 
the impact of BH feedback has a strong effect on a wide range of
halo mass, causing a reduction of the stellar mass fraction by  a
factor of 2 in $10^{11} \msun$ haloes mass and an order of magnitude in the 
high haloes mass at $10^{12} \msun$ relative to observations. The models are incompatible with the observational data across the full halo mass range probed.

We note that the effects of AGN feedback become significant at 
lower masses and are more severe than in the low redshift OWLS simulations, which agree well 
with observations \citep{McCarthy2010}, because our particle masses are lower by nearly two orders of 
magnitude (allowing us to convincingly resolve galaxies with stellar masses greater than $10^9\msun$ with 500 particles).
 
Comparing the simulations {\it BS09}  and {\it BS09-LE} , where the 
efficiency of feedback differs by an
order magnitude  ($\epsf =$ 0.15 and 0.015 respectively), it is clear
that the AGN feedback produces a similar suppression
of star formation and low stellar mass fractions in both runs, which
is consistent with the findings of BS09 and \citet{booth_schaye10}.

The tendency for the BHs to self-regulate, and thus to correlate strongly
with  the binding energy of the (inner) halo  has been emphasised by \cite{booth_schaye10}. As a 
result, a  strong correlation is expected  between the BH mass and $\mcrit$. 
The right panel of the Fig.~\ref{fig:effeed}  shows the BH mass as a function of $\mcrit$ for the 
 {\it BS09} and {\it BS09-LE} simulations. In order to provide an
observational baseline, black stars show estimates of the BH mass in 
48 galaxy-scale strong  gravitational lenses from the Sloan lens ACS
\citet{bandara2009} and grey lines represent the fit of 
data from  \citet{bandara2009} based on the $\mbh$--$\sigma_*$ relation
of \citet{gultekin2009}. The {\it BS09 } 
simulation  gives  a correlation between the BH mass and halo mass in good agreement 
with observations as found by \citet{booth_schaye10}. As expected, reducing the feedback efficiency by a factor 
of 10 produces  more massive BHs at fixed halo mass, leading to a $\mbh$--$\mcrit$ relation that is offset
from the observational data. As has already been shown by  \citet{booth_schaye10},
the total feedback energy remains the same in both models, consistent with the idea that the BHs
grow until they begin to unbind the gas halo of the system. At this point they become self-regulating growing in synchronize
with the halo mass.

The results from the two panels suggest that the self-regulated growth of the BH overwhelms the stellar mass growth of the central galaxy if
BH accretion is efficient. The gas,
which would have been able to form stars in the absence of AGN feedback, is
expelled from the system or prevented from accreting, starving the central galaxy of fuel for further star
formation. A comparison of the red and blue curves shows that the issue cannot be resolved by altering the
efficiency with which energy is deposited.  However,  following the scenario
discussed in section ~\ref{sec:newmodel}, we have already noted that a vital component of the 
accretion model is missing: gas that is supported by angular momentum cannot accrete as rapidly as the Bondi-like BS09
formula would suggest.

One unappealing solution is to only inject BHs into higher mass
haloes i.e. suppressing the impact of BHs on lower mass galaxies by hand. For example, injecting BHs at a halo mass of
$10^{11.5}\, h^{-1}\, \msun$ provides a good description of galaxy
properties. However, this is 
clearly unsatisfactory: at the resolution we consider, the haloes of $10^{10}\,h^{-1}\, \msun$ contain
more than 1500 particles and are well defined. There is no physical justification for not attempting to
model BHs in these haloes.

Another simple solution is to adopt a much lower accretion rate. Decreasing $\beta$ in Eq.~\ref{eq:beta} 
where the multiphase nature of the ISM is ignored, achieves the required goal of suppressing BH growth within galaxies. However, this strongly suppresses the accretion rates of early BHs making it difficult to form a BH population at high-redshift as required by observations
of massive BHs at redshifts $\sim 7$  \citep{mortlock11}. 

\subsection{Simulation with the AM accretion model} 
\begin{figure*} 
\begin{tabular}{cc}
\includegraphics[width=\columnwidth]{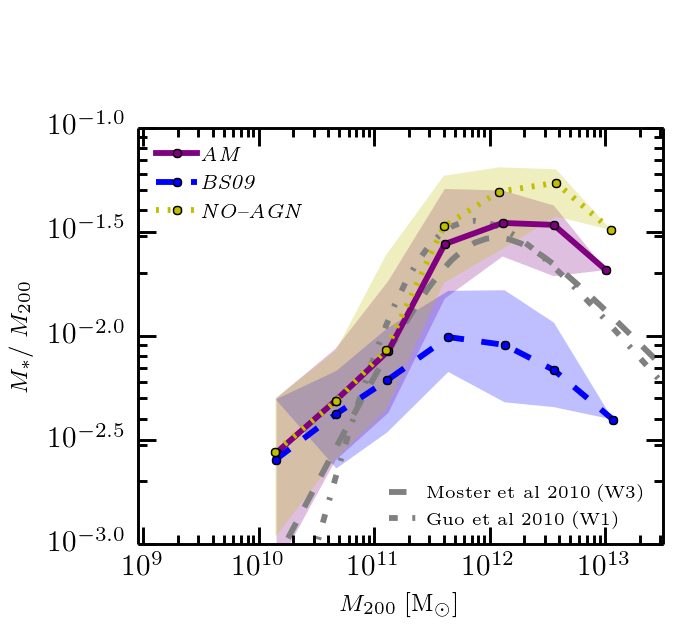} &
\includegraphics[width=\columnwidth]{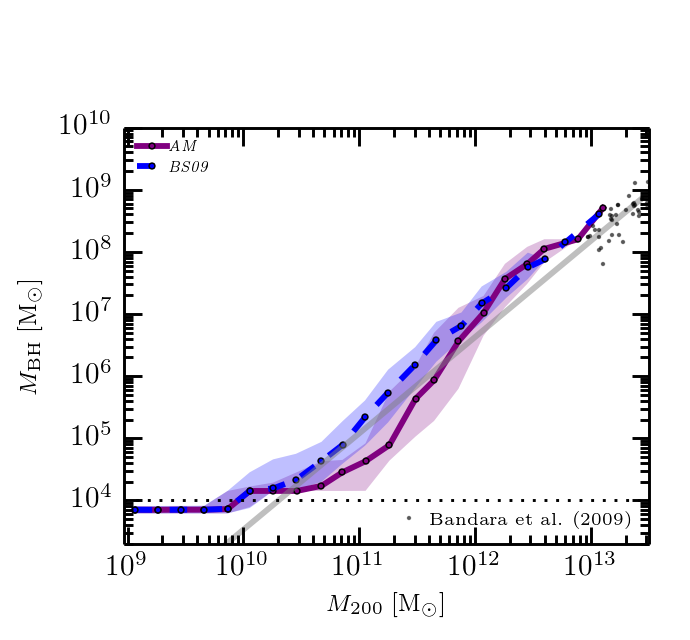} \\
\end{tabular} 
\caption{The left-hand panel shows the correlation of stellar mass
 fraction with $\mcrit$ for the angular momentum dependent accretion
 model, contrasted with two of models considered in section~\ref{subsec:bondimodels}.
  The right-hand panel shows the $\mbh$--$\mcrit$ relation for the same models. The yellow dotted line and blue (here dashed) 
coloured lines correspond to those in Fig.~\ref{fig:effeed}. The {\it AM} simulation is shown
in purple. The AM model results in resemble agreement 
with both the abundance matching relationship of the stellar mass fraction (grey lines) and with the 
observational $\mbh$-$\mcrit$ relation. The key to achieving
this match is that BHs residing in halos with mass below $\sim 10^{12}
\msun$ grow less than required for self-regulation.} 
\label{fig:simulationsnewmodels} 
\end{figure*}

In this section, we explore the effect of accounting for angular
momentum in the model. Fig.~\ref{fig:simulationsnewmodels}
is similar to Fig.~\ref{fig:effeed}, 
except that we show  the {\it AM} simulation where  the accretion 
rates account for angular momentum as  in Eq.~\ref{eq:newbhacr},
in purple. The left panel shows
that the {\it AM} simulation leads to stellar mass fractions that are
in much better agreement with the abundance matching relation
(grey lines) and still able to reproduce the turnover of the relation in high 
mass haloes. Note that the stellar mass fraction found in halos less
massive than $10^{12} \msun$   is similar to those in the {\it NO-AGN} simulation , supporting the idea 
that AGN feedback plays a minor role in regulating star formation in small galaxies.
Looking at the $\mbh$--$\mcrit$ relation in the right panel, we see
that the BHs residing in halos with masses lower than $10^{12}\, \msun$
in the {\it AM} simulation grow more slowly than expected for
self-regulation. However, above this halo mass, both models  lock onto the
same self-regulated relation, in good  agreement with observations.

\begin{figure}
\includegraphics[width=8.5cm,height=11.5cm,keepaspectratio]{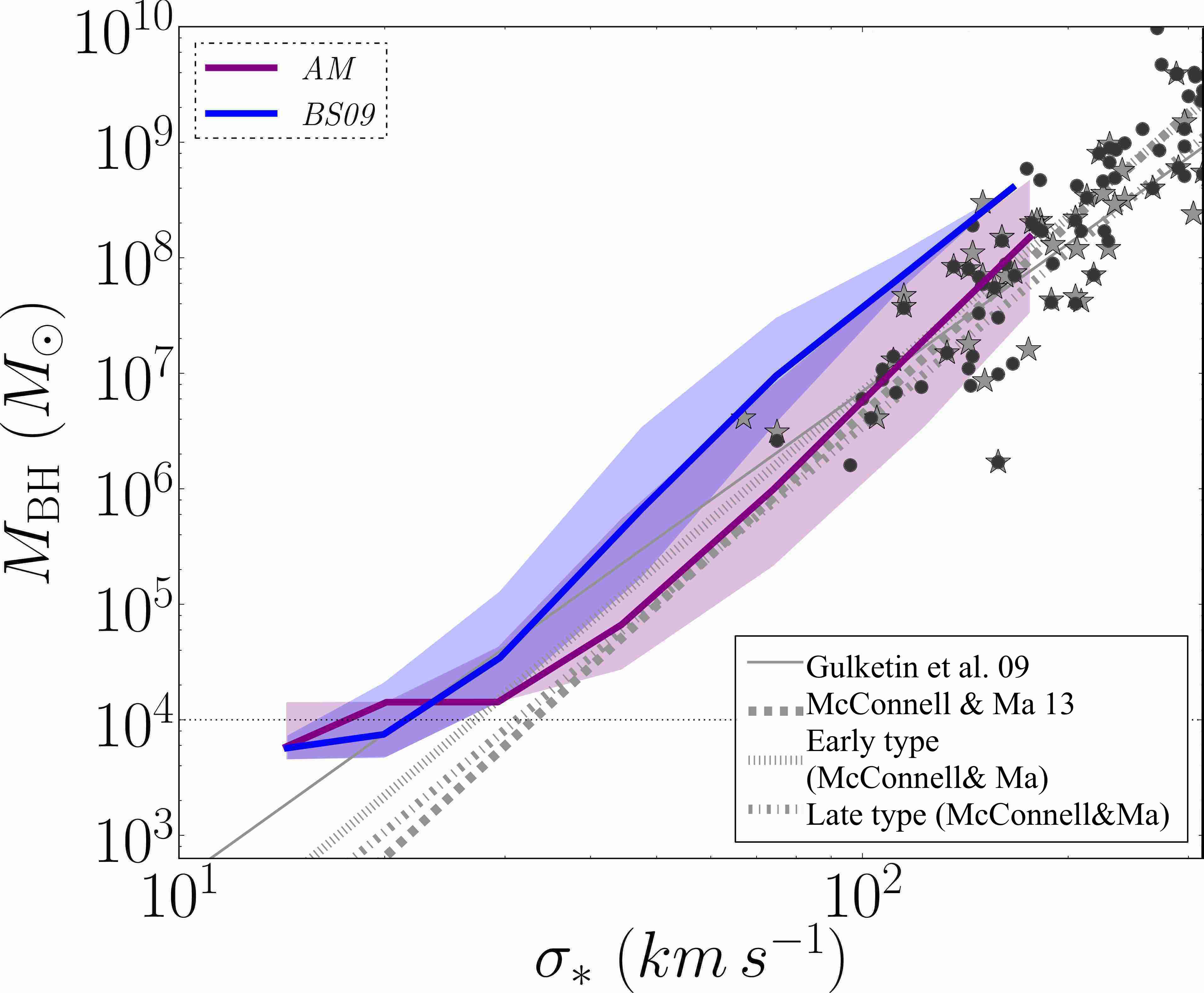}
\caption{ $\mbh$--$\sigma_{*}$ relation for the models presented in 
Fig.~\ref{fig:simulationsnewmodels}. Line colours are reproduced from that figure.
Observational data are shown in grey,  taken from
\citet{gultekin2009} and \citet{mcConnell_ma2013}  with 
the grey solid and dashed lines the authors' respective observational fits. The  grey dotted and 
dashed-dotted lines are fits only considering  early type galaxies  and late type galaxies respectively.  
The median relation from the {\it AM} simulation (purple line) is  
in good agreement with the fit expected  for the early type galaxies of \citet{mcConnell_ma2013} at the high $\sigma_*$ end 
while it encompasses the observational data for late type galaxies at the lower $\sigma_*$ values.} 
\label{fig:mbhmstar} 
\end{figure}

The model we have shown adopts $\Cvisc=2.1\times10^6$. The dependence 
of the outcome population on the value of this parameter is surprisingly weak --- as illustrated in Appendix \ref{appndx:variation_alphavisc}. We will argue below that the mass scale at which BH accretion becomes self-regulating is set by the properties of the halo rather than by the details of the AM accretion model.

While the BHs in lower mass haloes fall below the
(extrapolated) observational relation, the correlation between BH mass and halo mass is a very indirect observational
test of the models.  A better approach is to compare the BH mass to the stellar velocity dispersion
of the model. This is shown in Fig.~\ref{fig:mbhmstar}. 
Since the models have significantly different stellar mass and dynamics, we would expect the 
models to differ in this plot, and this is indeed seen. Observational data is shown 
from \citet*{gultekin2009} and \citet{mcConnell_ma2013} with the grey solid and dotted lines respectively showing  
the expected median relations for early-type galaxies 
(dotted line) and late-type galaxies (dot-dashed line). The difference between observational relations 
arises because McConnell \& Ma includes data for bright galaxy centrals (BCGs) of a cluster
resulting in a steeper fit. The {\it AM} model is clearly able  to pass the
test offered by  the $\mbh$-$\sigma_*$ relation.

\section{The Accretion History of Black Holes}
\label{sec:history_accretion}
\subsection{A Case Study}
\label{sub:history_accretion}

\begin{figure*} 
\includegraphics[width=2\columnwidth]{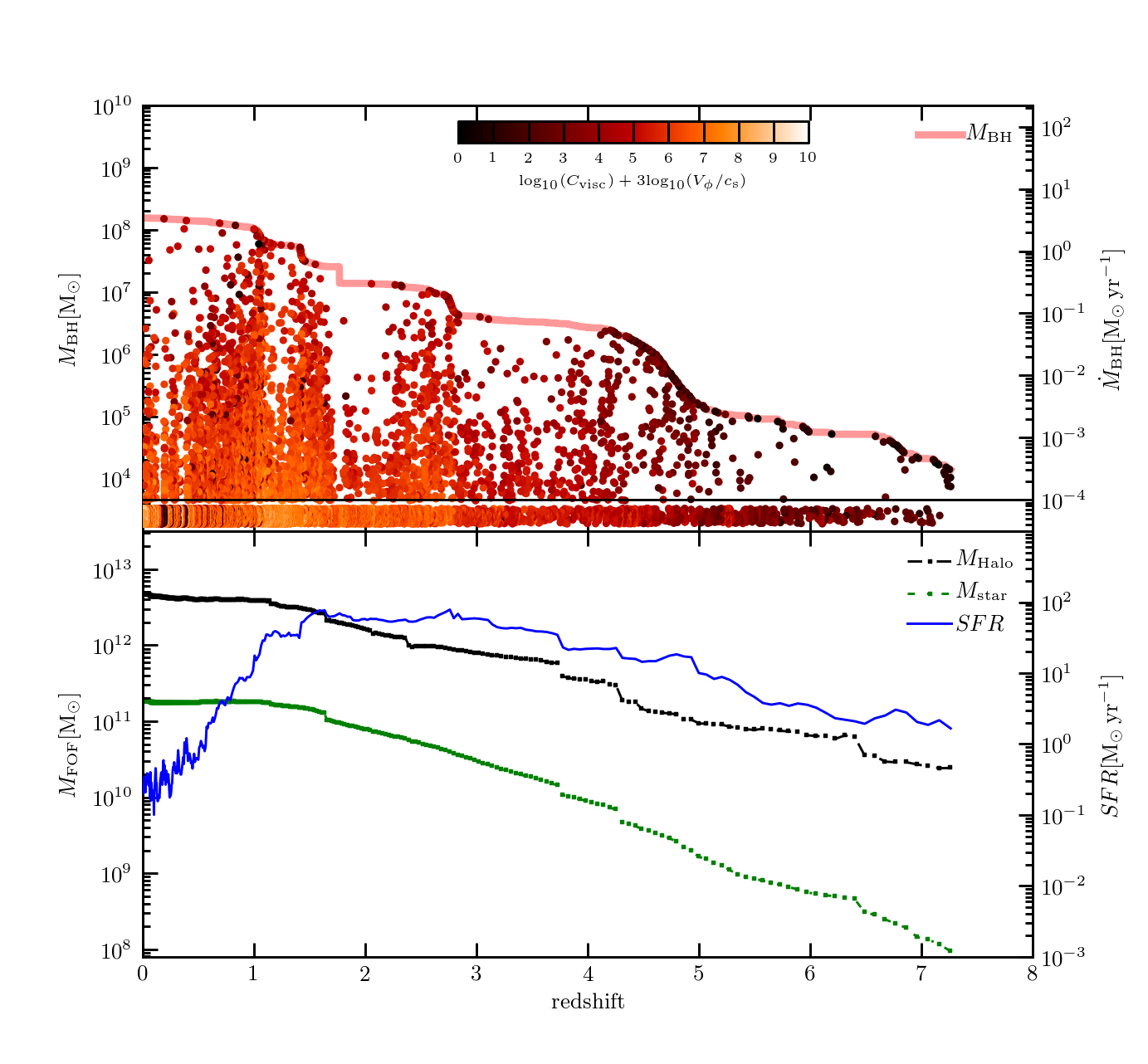}
\caption{History of a massive BH and its host galaxy in the
  {\it AM} simulation. The top panel shows the growth of the second most massive BH
  at $z=0$ as a function of redshift (red line and left-hand y-axis) as well
as its instantaneous mass accretion rate (coloured points and right-hand axis). The colour
code indicates the $\log$ suppression factor by which the mass accretion rate is reduced, 
$\log_{10}\Cvisc +3\times\log_{10}\Vphi/\cs$ where $\log_{10}\Cvisc = 6.3$ for the simulation. High accretion 
rates shows small suppression factor. The
red line represents simultaneously BH mass and Eddington accretion. Above $z=6$, the BH 
frequently accretes at Eddington limit,
but its mass and consequently its energy output are small. At intermediate redshift,
the BH undergoes phases of high mass accretion rate ( $z\sim 4.7$, $2.75$ and $1.1$), interspersed with periods 
of quiescence. Below $z=1$, the BH shows
frequent outbursts reaching 1--10\% of the Eddington rate. The bottom panel
shows the FOF mass (black long dashed-dotted line) and  total stellar mass
(green  dashed-dotted line) of the host halo (on the left y-axis) as a function of the
redshift. The blue solid line represents the total star formation rate
(right-hand y-axis).}
\label{fig:bhhistory} 
\end{figure*}

In order to understand the transition between the low- and high-mass halo behaviour better, we 
plot the ``timeline'' of a typical massive BH 
in the simulation in Fig.~\ref{fig:bhhistory}.  The curves in the upper panel
illustrate the mass of the most massive progenitor (red line,
left-hand axis), its accretion rate (coloured dots and right hand axis) and the Eddington
accretion rate (red line, right-hand axis) as a function of redshift. The points
plotted below the horizontal line represent BH accretion rates $\mdotbh < 10^{-4}\, \msun \yr^{-1}$. The colour code corresponds to
 $\log_{10}(\Cvisc)+3\log_{10}( \Vphi/\cs)$ (where $\log_{10} (\Cvisc)= 6.3$).
This quantity is the factor by which the mass accretion rate has been suppressed by taking AM into account. For 
comparison, the FOF mass and the total stellar mass of the host halo are shown in the lower panel as black and green lines
respectively (left-hand axis), along with the total star formation rate as the  blue line (right-hand axis).

The BH is injected with a seed mass of $1.4\times 10^4 \,\msun$.  Initially 
BH growth is limited by the Eddington accretion rate, but the BH growth soon settles down.   Although
the BH undergoes episodes of accretion close to the Eddington limit, the
absolute mass growth is small at high redshift. 
Over time, the suppression factor increases in importance. Eventually, promoted by a merger event that begins at $z\sim 5$, the BH 
undergoes a more sustained period of Eddington limited accretion. At the end of this event, the accretion
rate drops due to the suppression by the angular momentum. There is marginal further BH growth until $z\sim 3.0$ when another period of Eddington limited accretion
begins. The BH undergoes two further periods of Eddington limited accretion until
its host halo reaches a mass of $2\times10^{12}\msun$. It would seem 
reasonable to associate these high accretion rate events with QSO activity. During these high accretion rate periods, the suppression
factor is much lower than during the quiescent periods in between.   Below $z=1$, the nature of
accretion onto the BH appears to change again. Although the accretion rate fluctuates,
it rarely reaches the Eddington limit, typically peaking at  1\% to 10\%. The outburst are more frequent, but not sustained.
The correspondence between colour and accretion rate suggests 
that the accretion is limited by the angular momentum. The  suppression factor becomes more important as redshift decreases. 

The lower panel of Fig.~\ref{fig:bhhistory} allows us to contrast the
growth of the halo with that of the host galaxy of the BH. Since the plot
shows the total halo mass in each case, it does not identify galaxy mergers
well.  However, it does emphasise the steady growth of the halo and
its constituent galaxies, that is quite at odds with the episodic
growth of the BH. Despite the periods of strong BH activity,
the star formation rate of the halo seems little affected until the
halo reaches a mass of $2\times 10^{12} \msun$ at $z\sim2$.  In the following
outburst, the star formation rate is clearly set-back, with the 
outburst at $z\sim1.1$ initiating a sustained decay. 

\begin{figure*} 
\begin{center}
\includegraphics[width=20cm,height=20cm,keepaspectratio]{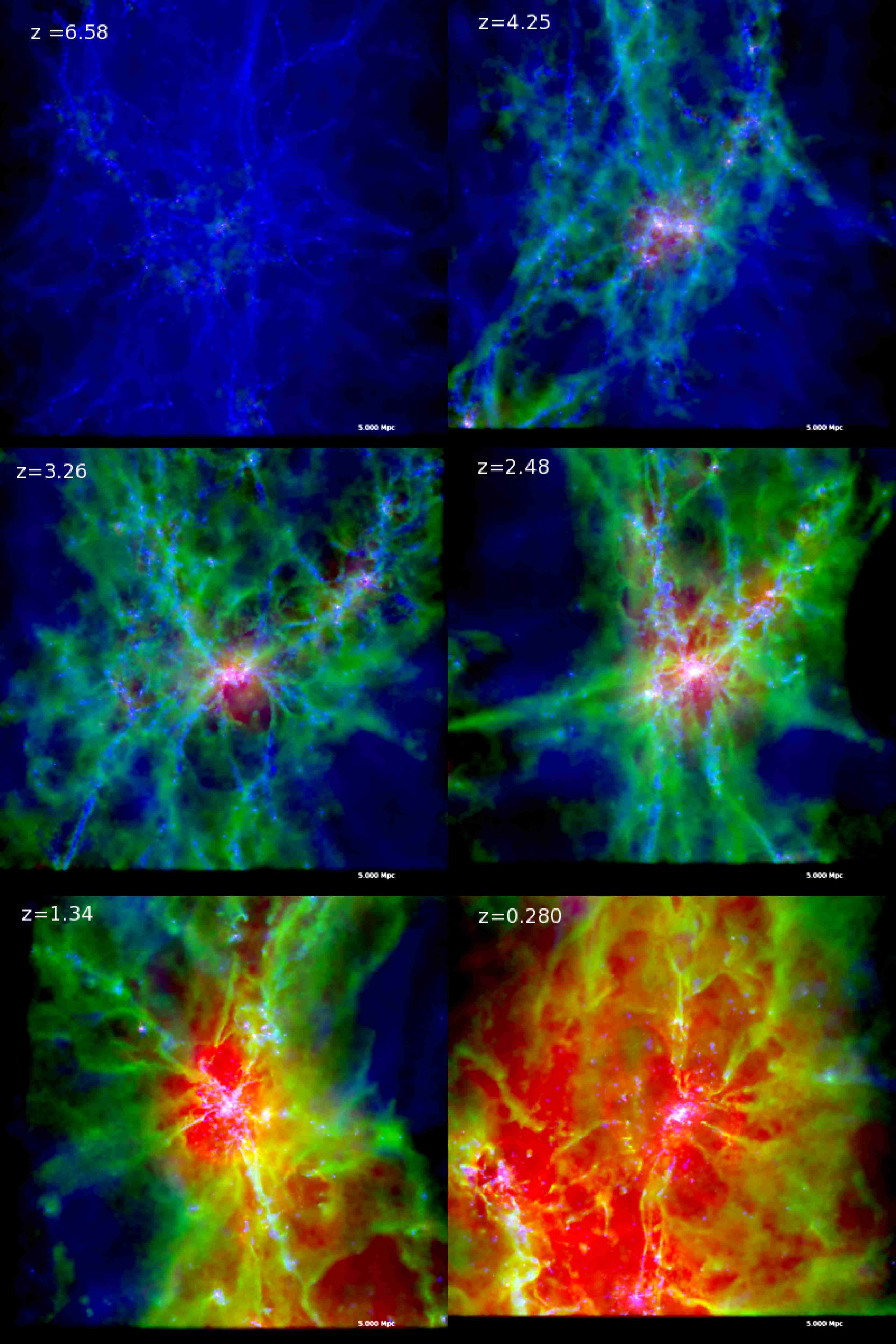}
\end{center}
\caption{This sequence of images illustrates the evolution of the 
galaxy and halo of the BH shown in ~Fig.\ref{fig:bhhistory}. The colour channels
show the projected gas density of gas with temperature $T<10^{4.5} \K$ (blue), 
$10^{4.5} < T<10^{5.5} \K$ (green) and  $T>10^{5.5} \K$ (red). The projected
star density is shown in pink. The box above shows 5 \Mpc  on each side,
representing $1/125$ of the total simulation volume.
The redshifts have been chosen to illustrate the change in the behaviour of AGN feedback.  The complete
movie can be downloaded  from the website shown in footnote 3.}
\label{fig:images} 
\end{figure*}

The simplest way to understand what drives the sustained accretion events seen
in Fig.~\ref{fig:bhhistory}, is to  examine a 
movie of the evolution (available from the website shown in footnote 3  
\footnote{\texttt{http://star--www.dur.ac.uk/\textasciitilde rgb/yrg2014/phase\_movie.avi}}).
Six frames from the movie are shown in Fig.~\ref{fig:images},
illustrating some of the important features. Examining the 
sequence in detail, however, shows that the outburst typically lags behind the 
obvious morphological disturbance. The first three outbursts follow a clear pattern. Immediately after the outburst, the accretion leads to the
expulsion of gas surrounding the BH. Although a strong wind is
generated, the void which this leaves around the BH is quickly
refilled by cool filaments that efficiently carry cool gas to the
centre of the halo. The final outburst is qualitatively different,
however: once the central gas is ejected, the surrounding matter does not flow in to replace it. Although
there continues to be some cool material at the centre of the halo,
this does not settle quickly around the BH.
We can understand the change in behaviour from the nature of accretion in haloes of different mass. As \citet{vandevoort2010}
have demonstrated in the OWLS simulations (see also \citealt{white_frenk91,keres2005,dekel_birnboim2006}), the 
fuelling of high mass haloes is qualitatively different from their low mass counterparts; high mass haloes are dominated by 
slow accretion from the hot halo, while lower mass haloes are supplied through rapid inflows along cold streams.

\subsection{BH accretion and its dependence on local gas
  properties}

\begin{figure*} 
\begin{tabular}{c}
\includegraphics[width=2\columnwidth]{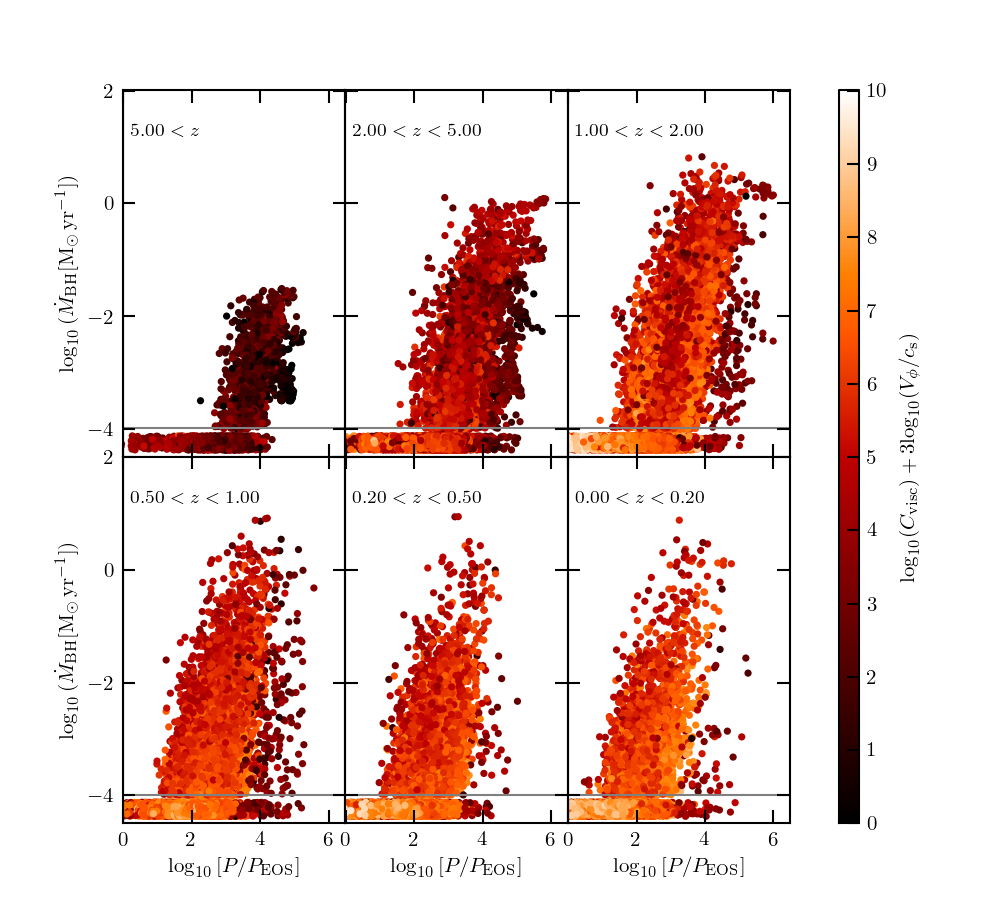}\\
\end{tabular} 
\caption{BH accretion histories plotted against the surrounding gas pressure  (in units of $P_{\rm EOS, 0}/k=2.3\times 10^3 {\rm cm}^{-3} \K$) for 
the most massive progenitor of the central BHs residing in the 9 halos with $\mcrit > 10^{12.5} \msun$ 
at z=0. Each panel represents a different redshift bin as indicated in all panels. The 
colour scale indicates  the  $\log_{10}$ suppression factor as in Fig.~\ref{fig:bhhistory}.  
Significant accretion events occur when the suppression factor is small and the surrounding gas pressure is high.}
\label{fig:mdot_pres_cb_vphi} 
\end{figure*}

In order to better understand the interplay between the pressure and angular momentum of the gas around the BH, we show 
mass accretion rates of the central BHs residing in the 9 most massive halos ($\mcrit>10^{12.5} \msun$
at $z=0$) in Fig.~\ref{fig:mdot_pres_cb_vphi}. 
Each panel divides the plot into different redshift bins showing BH accretion rate as a
function of the local effective fluid pressure, with points coloured by the $\log$ suppression factor as indicated in Fig.~\ref{fig:bhhistory}. 

The points plotted below the horizontal line represent BH accretion rates of $\mdotbh < 10^{-4} \,\msun \yr^{-1}$.
Points which are coloured dark red have a relatively low circular speed and low suppression factor;
orange through yellow and  white points, with higher circular speeds relative to local sound speed, are offset to much lower accretion rates. 
The diagram is somewhat complicated by points for which the accretion rate is Eddington limited,
but the Eddington limit only significantly distorts the diagram in the highest redshift panel. 

Fig.~\ref{fig:mdot_pres_cb_vphi} confirms the behaviour seen for a single BH in section~\ref{sub:history_accretion}. The accretion rates tend to 
increase strongly with the pressure of the surrounding gas. Most gas particles close to the BH
lie on the ISM equation of state so that the pressure and the sound
speed are closely correlated. In part, the trends in pressure arise because of the
strong modulation of the accretion rate by the BS09 $\alpha$ parameter in
Eqs.~\ref{eq:bh1} and \ref{eq:beta}. As the gas pressure increases in the simulation, the sub-grid model assumes that the density distribution of the ISM becomes increasingly clumpy and cold. This trend is modulated by the viscous suppression factor (shown by the colouring of the points)
since increasing pressures reduce the importance of the circulation speed (the increasing gas pressure reduces the Bondi 
radius with the result that the circulation radius and the viscous timescale are also reduced).
In addition to the main trend, a few points have a high pressure but
a low suppression factor and relatively low accretion rates. These are episodes when the BH accretes from the hot (but low density) intra-group medium, 
rather than from a central pool of ISM.

Focusing on the differences between the redshift panels, we see firstly that
the typical accretion rate increases across the 3 first panels, but
then gradually decreases. This confirms the trend seen for an individual BH in  Fig.~\ref{fig:bhhistory}. 
In the highest redshift panels, BHs accrete close to
their Eddington limits. In subsequent panels, the accretion
rates remain similar (and decline slightly) while the	 BHs grow
significantly in mass. Examination of Fig.~\ref{fig:bhhistory} shows that two factors contribute. Firstly, the
pressure  in the surrounding medium reaches a maximum
in the intermediate redshift panels and then slowly declines to lower
redshift. This possibly arises from the features of the BH's host galaxies: galaxies are less concentrated at lower redshifts as well as the galaxy gas fractions decline below $z=1$. As a result, gas presses around the BH less
tightly and tends not to accrete as strongly, despite the growth of
BH mass. For a fixed BH mass, the accretion rate depends on pressure as $P^{15/8}$ (for an effective equation of state with polytropic index 
$\gamma_{\rm EOS}=4/3$ and sub-grid density parameter $\beta=2$). This is indeed the slope
seen in the figure. The suppression factor also becomes more important at lower redshift
as clear from Fig.~\ref{fig:mdot_pres_cb_vphi}.

\begin{figure} 

\includegraphics[width=\columnwidth]{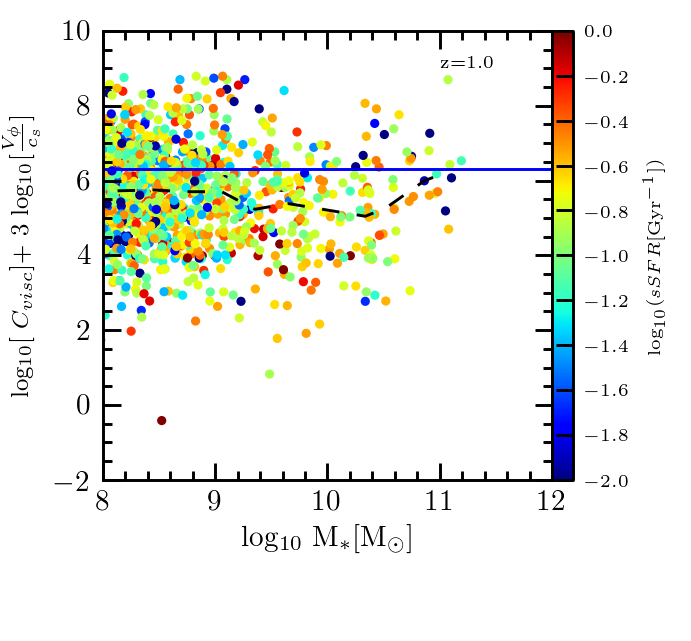}
\caption{The dependence of the accretion suppression factor, $\log_{10}[\Cvisc]+3 log_{10}[\Vphi /\cs]$, on galaxy stellar mass for the fiducial case, 
log$_{10} \Cvisc = 6.3$. The dashed line represents the median of the accretion suppression factor in bins of log
galaxy mass with a width of 0.3. 
The blue line shows the value of  log$_{10} \Cvisc$. The panel shows the relation at $z=1$, but the plot is similar at other redshifts.
In the majority of galaxies, $\Vphi$ and $\cs$ are comparable and the accretion rate is suppressed by a factor $\Cvisc$. 
In some galaxies, however, the suppression factor is much smaller and the angular momentum of the accreting gas is
no longer the limiting timescale. There is little trend in the median suppression factor with galaxy mass, as illustrated by the black dashed line,
or with specific star formation rate, as indicated by the colours of the points.}
\label{fig:csvphi_mstar} 
\end{figure}

In Fig.~\ref{fig:csvphi_mstar} we show the dependence of the accretion rate  suppression factor on galaxy stellar mass, with points coloured by the specific 
star formation rate (sSFR) of the host galaxy. The figure, shows that 
$\Vphi$ and $\cs$ are usually comparable (blue solid line), but that galaxies tend to under-go excursions to much smaller suppression factors, as we have seen in the 
case study. The black dashed line shows the median suppression factor in bins of log stellar mass with a width of 0.3 dex. It is clear that the suppression factor has only a weak 
dependence  on stellar mass with the accretion flow being only slightly less suppressed in the larger objects. This has important implications: the strong increase of 
BH mass as a function of halo mass is not the result of a direct dependence of $\Vphi$ on the halo mass. Instead, this dependence
must rise from the reaction of the halo to feedback from the BH, or from a halo mass dependence of the gas pressure around the 
BH.  We consider this further in section ~\ref{sec:discussion}.

\subsection{Implications for observable AGNs}

\begin{figure} 
\centering
\begin{tabular}{c}
\includegraphics[width=\columnwidth]{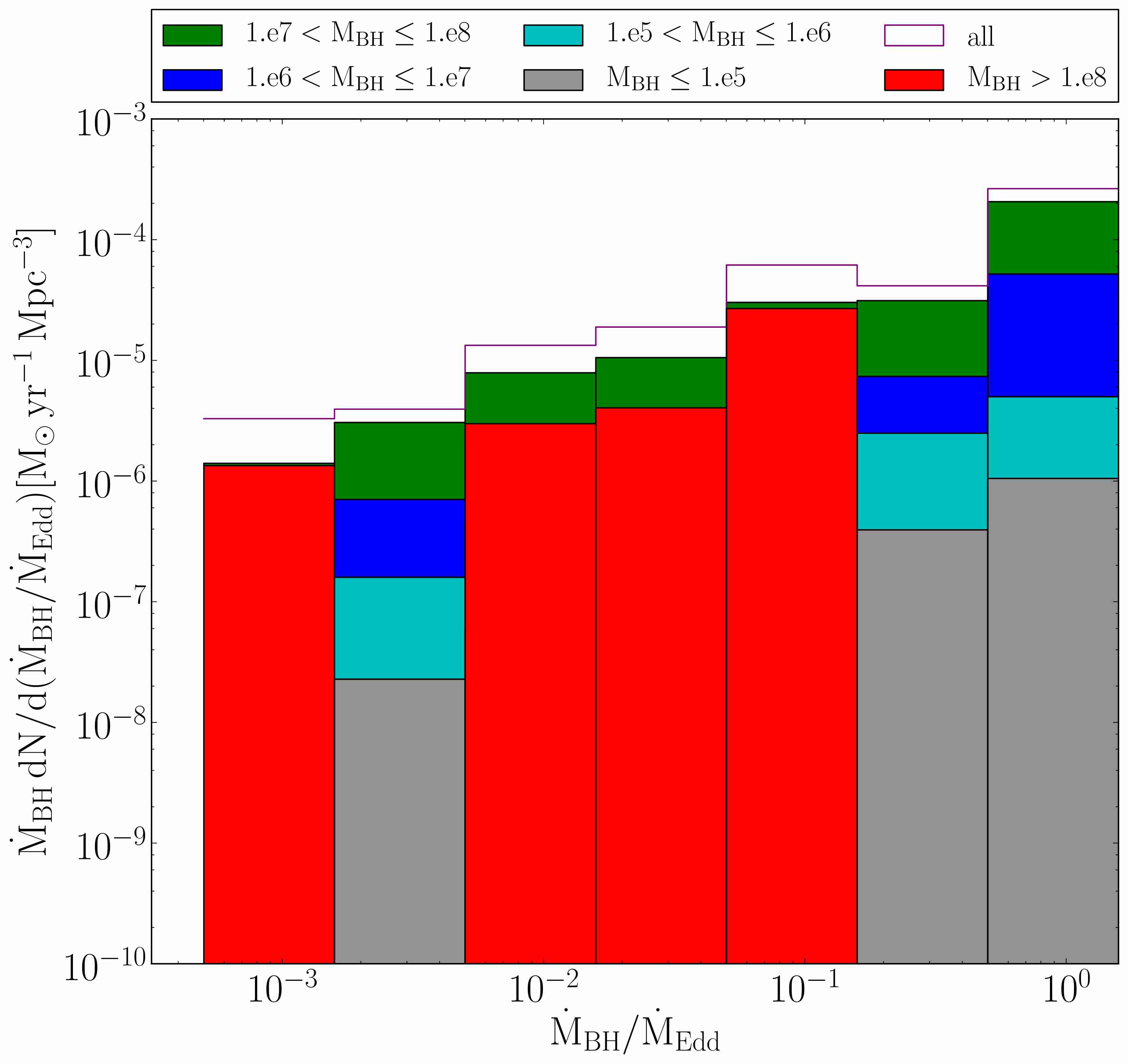} 
\end{tabular} 
\caption{The distribution of Eddington ratio weighted by the mass accretion rate 
of each accretion source (i.e. the accretion luminosity). This plot shows the distribution at $z=1$.
The total distribution is shown as a purple line; The contributions of BHs in different mass ranges are 
illustrated by different colours. For BHs less massive than $10^8\msun$, the  distributions
are similar, with larger BHs making a greater contribution to the BH mass budget. Above $10^8 \msun$, however,
high Eddington ratio sources become rare.}
\label{fig:histogram} 
\end{figure}

We will present an in-depth  comparison of the model with
observational data in a future paper. Here, we limit ourselves
to a brief look at the Eddington ratio distribution of accreting sources for the 
{\it AM } model. In Fig.~\ref{fig:histogram}, we plot a histogram of the Eddington ratio distribution (defined as $ \dot{M}_{\rm BH}/ \dot{M}_{Edd}$). 
We weight each source by its accretion luminosity so that the histogram shows the contribution to the growth of the BH mass 
density as a function of both BH mass and Eddington ratio.
We have chosen to plot the distribution at $z=1$ (combining outputs in the range $z=0.78$ to $1.12$); similar plots can be 
constructed at other redshifts, but the features of the distribution functions remain similar.

The purple solid line shows the Eddington ratio distribution of all BHs: the  integral under this plot 
gives the total mass growth rate of BHs at this epoch. Because, we have weighted the source
contribution by their accretion luminosity, the highest Eddington ratio sources
dominate the BH mass growth budget. If we had simply counted sources, the histogram would show that
low Eddington ratio sources are far more numerous. It is instructive,
however, to examine how this distribution is built up for BHs of different masses.  Different colour lines distinguish the
contributions from BHs of different masses, starting
from the lowest mass BHs, below $10^5 \msun$ (grey region), and building up to
show the contribution of the most massive BHs, above $10^8\msun$ (red region).
Because of the weighting by accretion rate (or equivalent to luminosity), the
smallest BHs contribute little to the total. The shape of the Eddington ratio distribution is similar for BH masses up
to $10^8\msun$.  The relative similarity of the distributions has previously been noted in 
observational data \citep{aird2012}.

Above a BH mass of $10^8 \msun$, however, high Eddington 
ratio sources become rare, and the growth of these BHs
 is dominated by accretion events that are $\sim 0.1$  of
Eddington. Although there are only a handful of such high mass BHs
in the simulation, they make a significant contribution to the
energy budget because of their mass. 
Studies of quasar clustering suggest that quasars do not occupy 
the most massive haloes at  low redshift. Instead, massive haloes are occupied by radio galaxies 
\citep{romanodiaz2010,angulo2012,fanidakis13},
whose power output tends to be in the form of a kinetic jet, but not in the form of highly visible
luminosity. It is quite plausible that this dichotomy arises from a change in the Eddington rate distribution
with BH mass \citep{meier2001,nemmen2006}.
We will test this aspect of the simulation against observations in more detail in a future paper using
a larger simulation volume.

\section{Discussion}

\label{sec:discussion}

The simulations that we have presented in the previous sections show the importance of  
accounting for the angular momentum of accreting gas. Compared to runs that 
implement a Bondi-like
accretion formula, accounting for the angular momentum of gas
surrounding the BH results in less BH growth and greater  stellar mass growth. This is particularly evident in haloes 
less massive than  $\sim 10^{11.5}\Msolar$. Compared to runs that do not include any 
feedback the revised BH accretion model leads to a sharp turn-over in the stellar mass fraction of haloes.

In this section, we briefly consider the physical processes that establish this behaviour. In particular, we would like to 
understand what creates the transition in BH accretion efficiency at $10^{11} - 10^{12}\Msolar$. It is notable that 
this mass scale is weakly dependent on the accretion viscosity parameter 
$C_{\rm visc}$ (see appendix \ref{appndx:variation_alphavisc}) which suggests than this mass scale is not caused by the subgrid accretion model. Moreover, this mass scale approximately corresponds to the transition between
``rapid cooling'' haloes, in which the cooling time is short compared to the dynamical time, and ``hydrostatic'' haloes in which
the cooling time is long compared to the halo dynamical time. The key to understanding the nature of this transition 
is to examine the balance of gas accretion versus star formation and outflows. Since the mass
locked up in stars is relatively small, the most important consideration is the outflow of gas. This may be driven by stellar feedback or AGN feedback from the growth of the BH. Recent papers have emphasised that outflows lead to
self-regulation, both for feedback from star formation and from BHs
(\citealt{white_frenk91,bower2006,bower2008} also \citealt{scha10,booth_schaye10,dubois2013} and \citealt{puchwein13}). 
The observed dependence of stellar mass fraction on halo mass (and hence the shape of the galaxy mass function) requires that
star formation dominates the self-regulation at low halo masses, while it is
dominated by BHs at high stellar masses.  The challenge is to
understand why this switchover occurs.

We begin by considering the physical behaviour of Bondi-like accretion models. In the simulations presented here, we use a lower 
particle mass than BS09 to better sample star formation in lower mass galaxies, but the same ISM equation of state.  
Without accounting for the angular momentum, BH accretion is efficient across a wide range halo mass 
\citep{booth_schaye09,booth_schaye10}. This results in the balance of gas inflow and outflow being regulated by BH 
growth at the expense of the growth of stellar mass. 
This drives the stellar mass function to a power-law shape that is in complete contrast to the
observed Schechter function (see \citealt{benson2003}). The balance of BH and stellar mass growth
cannot be redressed by altering the feedback parameters (as opposed to accretion model): reducing the efficiency of 
BH {\it feedback} results in greater BH growth but does not improve the efficiency 
with which inflowing material is converted into stars.

In contrast, including angular momentum in the calculation reduces BH accretion rates in smaller haloes
allowing the inflow and outflow in these haloes to be regulated by star formation. Above a halo mass of 
$10^{11.5}\Msolar$, however, BH accretion becomes more efficient, creating a turnover in the 
stellar mass -- halo mass relation.

We began this paper by stressing the role that AGN feedback plays in shaping the properties of galaxies. In semi-analytic 
models, the key to reproducing the observed dependence of the stellar mass fraction on halo mass is the link between effective 
AGN feedback and the halo cooling time. This results in the full impact of AGN 
accretion only being felt in haloes more massive than $\sim 10^{11.5}\msun$. We have shown that the angular momentum dependent
BH accretion model used in this paper has a similar overall impact, limiting BH self regulation to the most massive haloes in the
simulation.

 But what sets the scale at which this occurs?  One possibility is that the suppression factor $\Vphi/\cs$ may be strongly dependent on 
system mass. However, we demonstrate that this is {\it not} the case in Fig.~\ref{fig:csvphi_mstar}. The absence of a strong correlation
in this plot shows that the rapid rise of BH mass with halo mass does not directly result from the dependence of the accretion 
rate on $\Vphi$.  The origin of this halo mass dependence is more subtle, and we consider other possibilities below.

We have already noted that the halo mass scale corresponds to the transition from
the rapid cooling regime (in which accretion shocks cool rapidly) to the hydrostatic regime (in which the cooling time of gas 
is long compared to the dynamical time of the system). We can expect that this creates a difference in the mode
by which gas is supplied to the central galaxy \citep{white_frenk91,keres2005,dekel_birnboim2006,vandevoort2010}: in lower mass haloes, accreted gas flows almost directly onto the galaxy, fuelling a cold gas disk at the centre of the halo, while in more massive systems accreted gas will be shock heated to the virial temperature of the halo before
slowly cooling out at the centre.  These differences are evident in our simulations, but this cannot be the 
complete picture.  The different modes of halo accretion may well explain why AGN feedback is much more effective at suppressing
star formation in  the central galaxies of hydrostatic haloes, but it does not explain why BHs accrete slowly in 
low-mass haloes, while they are able to self-regulate in massive haloes.

\begin{figure} 
\centering
\begin{tabular}{c}
\includegraphics[width=\columnwidth]{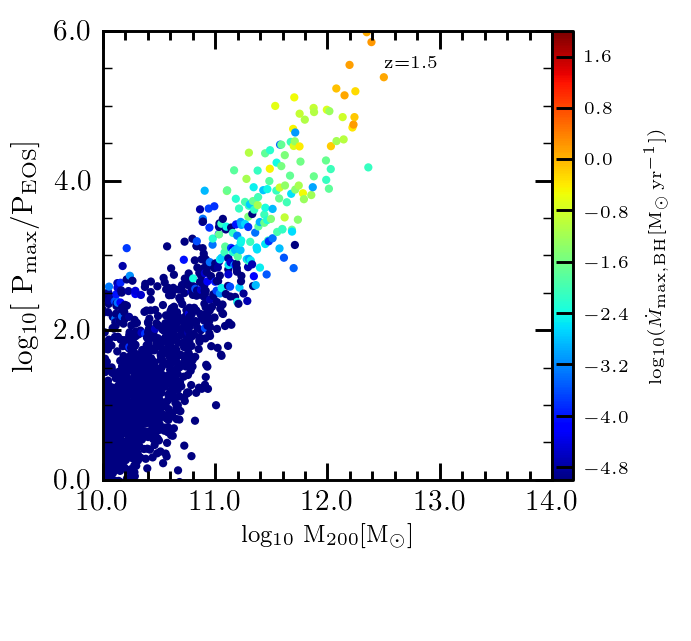} 
\end{tabular} 
\caption{The pressure of gas surrounding the BH as a function of the halo mass. As can be seen from
Fig.~\ref{fig:csvphi_mstar}, there are large fluctuations in pressure between each snapshot output.  In order to make the trend, 
clear we plot the maximum pressure experienced by each BH in the redshift range 1.35-1.65. The pressure is given in terms
of the equation of state threshold pressure $P_{\rm EOS,0}$. Points are coloured by the maximum BH accretion rate in the 
same redshift interval.  As halo mass increases, the effective pressure of the 
gas surrounding the BH increases rapidly leading to greatly increased BH accretion rates.
} 
\label{fig:pres_mhalo} 
\end{figure}

There are at least 5 mechanisms that might couple the accretion efficiency of low mass objects to the mode of gas accretion and the structure of the surrounding halo. We consider each briefly below.
\begin{enumerate}
\item A transition in the angular momentum of the accreted material. Direct accretion of cool filaments may result in greater angular momentum in the disks formed in lower mass haloes.  Such a transition could be driven by the change from rapid cooling (with accretion occurring through misaligned streams) to more spherical accretion through the surrounding hot gas halo.
 Examining Fig.~\ref{fig:csvphi_mstar}, however, shows that the halo mass dependence of this factor is extremely weak.

\item A transition in the merger rate. Efficient accretion is likely to be driven by mergers of stellar clumps that disrupt the angular momentum of the gas disk.  While it is not immediately evident why this would be strongly dependent on the halo mass (since the halo merger rate is weakly dependent on halo mass), and indeed this explanation is also ruled 
out by Fig.~\ref{fig:csvphi_mstar}.

\item A transition in the impact of the gas heated by the BH feedback. In low mass haloes, material heated by the BH may simply leave the system without redistributing its energy effectively. In contrast, the energy injected into higher mass haloes may be effectively trapped and shared with the surrounding particles. In Fig.~\ref{fig:images}, we can clearly identify expanding shells of material in the massive haloes; however, when we trace the final location of particles as a function of halo mass (relative to the halo virial radius), no clear trend with halo mass is evident. 

\item The ability of the halo to recover after an AGN feedback event. In lower mass haloes, the cooling time of inflowing material is short, and its filamentary geometry tends to limit the effect of a feedback outburst. Examination of individual events suggests that, in lower mass haloes, the outburst only affects the  material in the galaxy briefly, and that the system is quickly refuelled by accretion from the surrounding cold streams. Accretion in larger haloes is more fragile, since the cooling time is long and the cool streams evaporate before reaching the central object: once the streams are disrupted by an AGN outburst, they struggle to re-establish themselves \citep{dubois2013}. While this scenario explains the absence of star formation in the most massive galaxies, it does not account for the 
relatively weak growth of BHs in smaller haloes.

\item The pressure dependence of the BH accretion rate. Fig.~\ref{fig:mdot_pres_cb_vphi}  shows that the accretion rate of the BH depends
very strongly on the effective pressure of the surrounding gas. In Fig.~\ref{fig:pres_mhalo}, we show the halo mass dependence of the pressure. In order to smooth
over fluctuations in the accretion rate, we plot the maximum accretion rate reached in the redshift interval $z=1.35$--$1.65$ and colour points by the maximum
BH accretion rate. A strong correlation
can be seen, such that in lower mass haloes, the BH is surrounded by lower pressure gas. This leads to significantly lower BH mass accretion rates.
This trend re-enforces itself: the lower accretion rates lead to slower BH growth and hence to a lower accretion rate at a fixed gas pressure. In contrast, 
higher mass haloes generate higher pressures around the BH, leading to higher accretion rates and higher BH masses which, in turn, have 
higher accretion rates at a fixed pressure.
\end{enumerate}

Qualitatively, this picture is confirmed by the  images shown in Fig.~\ref{fig:images}, and in the linked animation (see footnote 3).
During the early universe, haloes are able to recover quickly due to the strong filamentary nature of their accretion. Although there are sporadic 
bursts of energy injection from the BH, these do little to suppress the continuous inflow of gas filaments. At late times, however, the BH is
able to couple effectively to the surrounding halo, disrupting filamentary gas accretion and establishing a hot gas envelope that appears to be 
pressure supported. The disruption of filaments is only part of the BH impact, however. While it may explain the suppression of gas inflow
(and hence star formation) in high mass haloes, it does not provide an explanation for the low accretion rates of BHs in low mass haloes. 
The low growth rates of BHs in lower mass haloes can, however, be accounted for by looking at the pressure of the gas surrounding the 
BH.  The pressure shows a strong increase with halo mass, greatly increasing the resulting BH accretion rates.

Fig.~\ref{fig:csvphi_mstar} shows that the suppression of the accretion rate due to the angular momentum does not have a strong halo mass 
dependence. Over long timescales, and averaged over the population of objects, the effect is to renormalise the accretion
rate relative to the Bondi-like BS09 formula~\ref{eq:bh1}. The angular momentum model does however, play an important role
in the determining the duty cycle of energy input. As a result of the angular momentum variations, 
high $\mdotbh$ accretion events are interspersed with quiescent periods during which the stellar component of the galaxy grows strongly, as 
shown in Fig.~\ref{fig:bhhistory}. The history of each individual BH is, however, complex and a combination of all the mechanisms may contribute at some level.
The net effect is similar, but not identical, to the mechanisms invoked by semi-analytic models \citep{bower2006, croton2006}; significantly
more work is required to explore the similarities and differences, and to capture the behaviour of the simulation
in a simple but quantitative model.
          
\section{Conclusions}
\label{sec:conclusions}

We have presented a development of the sub-grid  accretion model of BS09, based on Springel et al. 2005, that
takes into account the angular momentum of the accreting gas using the local SPH kernel
to estimate the circular speed at the Bondi radius and to define the circularisation
radius of the material passing through the Bondi radius. This creates an additional timescale in the accretion of gas onto a BH which 
characterises the transport of material through the disk and the fraction of this material
ejected out of the disk before it is accreted by the BH. We incorporate this new timescale into a revised accretion rate estimate
that can be simply implemented as a sub-grid model in cosmological simulations.

Two BH accretion models are tested: one using the prescription of BS09 and one with the updated angular momentum model. The
cosmological simulations include gas cooling, metal enrichment, star
formation and supernova feedback which we keep fixed. We show that simulations that 
do not account for the circulation of gas in the neighbourhood of the BH result in
self-regulation of the gas supply by BH growth rather than star
formation. As a result BH masses correlate well with the parent halo
over a wide range of scales, but self-regulation by the BH
results in insufficient star formation to match the observed dependence of
stellar mass on halo mass in low mass halos.

In contrast, when the angular momentum is taken into account, the accretion time scale is increased 
and BH accretion rates are strongly suppressed in haloes less massive than $10^{11.5}\msun$. This allows stars to form efficiently and for 
the balance of gas inflow and outflow to be set by stellar feedback. At higher halo mass, the effective pressure of gas surrounding the 
BH increases, compensating for the lower accretion rates of the angular momentum model. As a result, a strong division into two 
regimes of galaxy formation emerges from the model. Below a halo mass of $\sim 10^{11.5}\msun$, we find that the presence of the BH has little effect on 
forming galaxies, but above this halo mass star formation is suppressed and galaxies 
grow much more slowly in stellar mass compared to the growth of their haloes. The model broadly matches the observed stellar mass fractions of
haloes  and reproduces the expected correlation between the stellar velocity dispersion and BH mass \citep{mcConnell_ma2013}. 
The distribution of BH accretion rates also seems compatible with that observed \citep{aird2012}. The model thus provides a promising
prescription for cosmological simulations.

We speculate that there are two critical factors that establish the break in the galaxy stellar mass function. The first is the ability of the BH to accrete the material; accounting for the angular momentum of the material surrounding the BH suppresses accretion from quiescent disks allowing the growth of the galaxy in galactic mass haloes to be regulated by star formation.  The  second is the response of the halo gas to outbursts of BH activity. This introduces a strong dependence on the ratio of cooling time to dynamical time of the halo through (i) the ability of the inflowing filaments to re-established themselves following a brief episode of energy injection from the BH and (ii) the pressurisation of the halo as a quasi-hydrostatic hot corona is established.  In lower mass haloes the cold gas disk surrounding the BH quickly recovers because of the rapid supply of fresh material from the surrounding cosmic web. In high mass haloes, filamentary accretion is disrupted by the hot corona. As a result, star formation in the most massive galaxies is strongly 
suppressed and the break in the stellar mass function is established.

The BH model presented here forms the basis of the accretion model used
in the EAGLE simulation project \citep{schaye2015}. The EAGLE simulations use
$\beta=0$, which allows the $\alpha$ factor, discussed in
section~\ref{sec:owlsmodel}, to be absorbed into the parameter $\Cvisc$. The
EAGLE simulations have larger volume than the work presented here allowing an 
examination of the observable predictions of the model, such as the evolution of
the accretion rate distribution and the correlations between episodes of star
formation and BH growth in more detail. 

Although the current simulations suggest a promising way forward to include the angular momentum 
of infalling gas in sub-grid models of BH accretion, a great deal of work remains to be done. 
Larger volume calculations are required
to determine accurately the break in the galaxy stellar mass function and to
probe the quasar luminosity function, since these are both determined by rare
objects. Clearly, one of the most important future steps is to base the choice of the $\Cvisc$ parameter on finer-scale simulations that simultaneously
resolve the circularisation radius and the multiphase structure and turbulence of the interstellar medium around the BH. For the foreseeable future, such a multiscale approach seems the only feasible route to capture simultaneously the microphysics of BH accretion and the large-scale distribution of galaxies and quasars.

\section*{Acknowledgements}
We thank James Mullaney, Nicolas Tejos, James Aird and Rob Thacker for a careful reading of the paper, useful comments and discussions.  This work would have not be possible without Lydia Heck's technical support. YRG gratefully acknowledges financial support from the Mexican Council for Science and Technology (CONACYT) (Studentship No. 213183). CSF acknowledges an ERC Advanced Investigator grant, COSMIWAY (GA 267291). This work used the DiRAC Data Centric system at Durham University, operated by the Institute for Computational Cosmology on behalf of the STFC DiRAC HPC Facility (www.dirac.ac.uk). This equipment was funded by BIS National E-infrastructure capital grant ST/K00042X/1, STFC capital grant
ST/H008519/1, and STFC DiRAC Operations grant ST/K003267/1 and Durham
University. DiRAC is part of the National E-Infrastructure. This work was supported by the Science and Technology Facilities Council (ST/F001166/1,ST/L00075X/1), European Research Council under the European Union's Seventh Framework Programme (FP7/2007-2013)/ERC Grant agreement
278594-GasAroundGalaxies, the Netherlands Organisation for Scientific Research (NWO)  and Interuniversity Attraction Poles Programme initiated by the Belgian Science Policy Office ([AP P7/08 CHARM]. We thank contributors to SciPy \footnote{http://www.scipy.org} , Matplotlib \footnote{http://www.matplotlib.sourceforge.net} , and
the Python programming language \footnote{http://www.python.org}.

\appendix

\section{Convergence Tests}

\label{appndx:convergencetests}
\subsection{Convergence of $\Vphi$ in Idealised Galaxies}

\label{appndx:idealised_galaxies}

\begin{table*}
\caption{A list of the simulations for an idealised disk galaxy. The simulations have identical initial conditions, differing only in the number of particles used to realise the galaxy. }
\begin{tabular}{|l|c|c|c|}
\hline
Name& particle mass&  $\sigma_{\Vphi}/\hbox{km}\hbox{s}^{-1}$& $\sigma_{\cs}/\hbox{km}\hbox{s}^{-1}$\\
\hline
{\it IDEAL-EAGLE-RES}    &   $1.4\times 10^6$       &  1.94   &  2.41 \\                
\\
{\it IDEAL-EAGLE-RES5}  &    $2.5\times 10^5$      &  1.87   &  2.73 \\
\\
{\it IDEAL-EAGLE-RES25}  &  $5.6 \times 10^4$     &   1.63  &  4.10 \\
\\
\hline
\end{tabular}
\label{table:idealised_galaxy}
\end{table*}

\begin{figure*} 

\begin{tabular}{cc}
\includegraphics[width=\columnwidth]{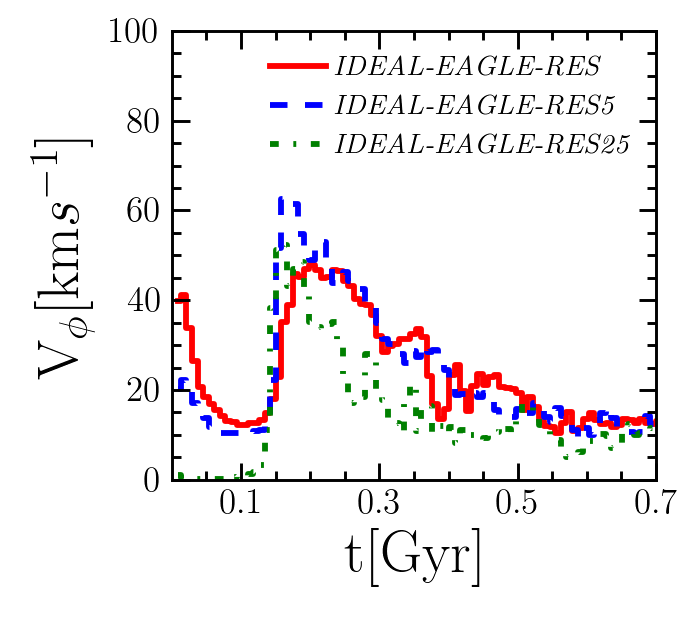}
&
\includegraphics[width=\columnwidth]{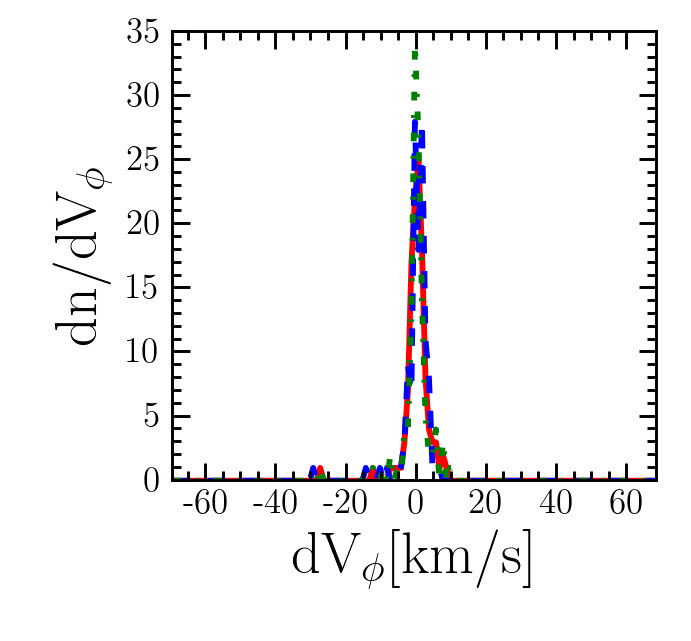}
\\
\includegraphics[width=\columnwidth]{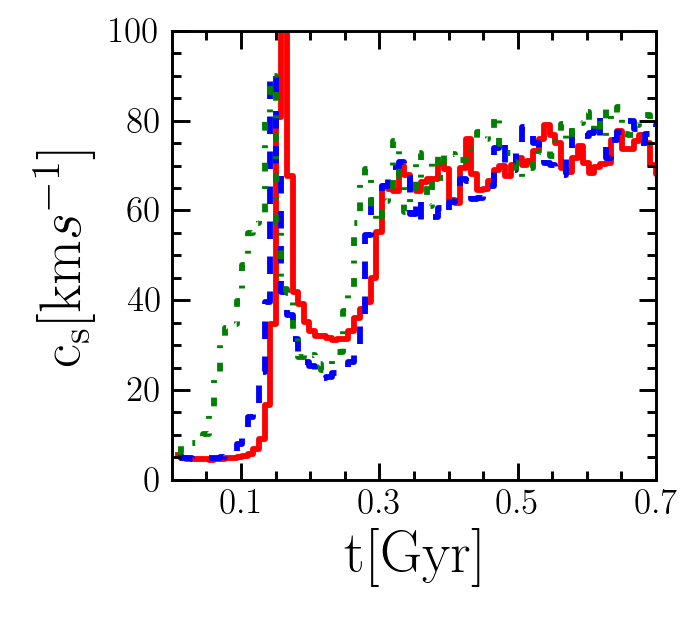}
&
\includegraphics[width=\columnwidth]{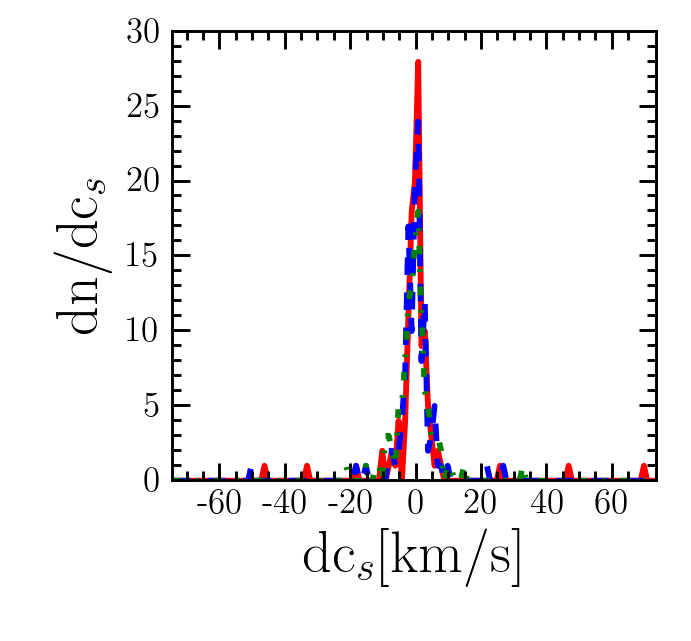}
\\
\end{tabular} 
\caption{The evolution of $\Vphi$ and $\cs$ in an idealised galaxy simulation carried out with increasing particle number.
The simulations do not include cooling or feedback and all assume the same gravitational softening length.
 {\it The left-hand panels} shows  the evolution of $\Vphi$ and $\cs$ as a function of time.
The galaxies quickly relax into a stable configuration in which the circulation speed measured within the BH 
smoothing length stabilises to values around 10  km s$^{-1}$, while $\cs$ converges to values around 80 km s$^{-1}$.  Despite the 
differences in particle number, similar values are found in each of the simulations.
{\it The right-hand panels } show the distribution of change $\Vphi$ and $\cs$ between
consecutive timesteps for different resolutions. The distributions show little dependence on particle number.}
\label{fig:idealised_galaxy} 
\end{figure*}

An important component of our approach to modelling BH accretion is to estimate the circulation speed
of gas at the Bondi radius using the curl of the velocity field measured from an SPH kernel centered on the BH.
In this section we consider the convergence of $\Vphi$ in idealised galaxies as the number of particles is increased.  It should be noted that 
measurements of the curl of the velocity field are commonly used in SPH simulations to suppress excessive viscosity in 
rotating disks  (e.g. \citealt{balsara1996}), and that this aspect of our model is not particularly novel.

In order to investigate the convergence of the estimate of $\Vphi$, we required well controlled simulations in which we are able to minimise differences arising 
in other sub-grid physics components such as the star formation implementation, cooling and the treatment of the ISM. The difficulty of obtaining
meaningful convergence results in the presence of sub-grid physics is discussed in depth in \cite*{schaye2015}.
We set up  a suite of simulations of idealised  disk galaxies generated  from the same initial conditions as used by \cite{springel2005, dallavecchia_schaye08} and \cite{stringer12}. 
These papers showed that even such idealised galaxies are very sensitive to resolution for identical physical assumptions.  In order to minimise these issues, 
we disabled cooling and feedback from SN and AGN. Our aim is to undertake the same calculation more accurately as we increase the number of particles, 
and we therefore use the same gravitational softening length in each run. Even under these conditions, convergence is not guaranteed since the amplification
noise in the spiral density waves will not be identical (e.g. \citealt{sellwood12}).

The idealised simulations consist of an static dark matter halo of mass  10$^{12}h^{-1}\msun$ with a Hernquist mass distribution \citep{hernquist1990} and an exponential disk of stars and gas.  The halo has a dimensionless spin parameter $\lambda=0.33$. The disk contains 4 percent of both the total mass and the total angular momentum and the bulge contains 0.014 percent of the total mass. The initial gas fraction of the disk is 0.3 and the rest consists of stars. The disk and bulge  scale heights are set to 10 percent of the radial disk scale lengths
and the vertical gas distribution is estimated to be in hydrostatic equilibrium using an iterative procedure.  The initial masses of  particles in the {\it IDEAL-EAGLE-RES} simulation 
were set to be equal to the value of the cosmological simulations.  The simulation was then repeated with increasing particle number, as shown in Table \ref{table:idealised_galaxy}. 
We follow the galaxies' evolution for 700~Myr, by which time the initial transients have decayed. 

We focus on the effects of resolution on the evolution of $\Vphi$ and $\cs$.  The left panel of  Fig.~\ref{fig:idealised_galaxy} shows the evolution of $\Vphi$ (top) and $\cs$ (bottom). 
During the initial transients, $\Vphi$ reaches high values of $\sim$ 60  km s$^{-1}$ and $\cs$ of $\sim$ 100 km s$^{-1}$  more or less independent on resolution. 
As the disk relaxes, values of $\Vphi$ become much smaller, while $\cs$ remains high. After 500 Myr, the measured values of $\Vphi$ and $\cs$ are almost independent of resolution.
We use these simulations to investigate the noise in the measurement of $\Vphi$ by examining the fluctuations in $\Vphi$ (and $\cs$). In the absence of violent
feedback events we expect $\Vphi$ to vary smoothly between time steps. The distribution of $\Vphi$ for each simulations is shown in the upper 
right panel of Fig.~\ref{fig:idealised_galaxy} (the distribution of variations in $\cs$ are shown in the lower panel). In order to avoid undue influence from 
outliers, we quantify the width of the distribution as half of the 16$^{th}$ to 84$^{th}$ percentile range (for a Gaussian distribution is equivalent to the standard deviation). 
Values are given in Table~\ref{table:idealised_galaxy}. The distribution has a similar width
regardless of the particle number.

\subsection{Convergence of the galaxy formation model}

Because of the small relative scales of the BH and its accretion
disk relative to the scale of the simulation, the most
challenging aspect of the calculation is to
adequately describe the properties of the multi-phase gas in the region 
around the BH.  Because of this, it would not be surprising if the
parameters of the model were functions of scale. Nevertheless, we
present a comparison of two versions of  simulation {\it  AM }  carried out at different 
resolutions in the standard 25 Mpc volume. Specifically, we run the simulation {\it AM LOWRES} 
with a lower  mass resolution by a factor of 8 (with $2\times180^3$ particles and with mass $5.88 \times 10^7 \,  h^{-1}\msun$
and  $1.18\times 10^7\,  h^{-1} \msun$ for dark matter and gas particles respectively).

\begin{figure*} 
\begin{tabular}{cc}
\includegraphics[width=\columnwidth]{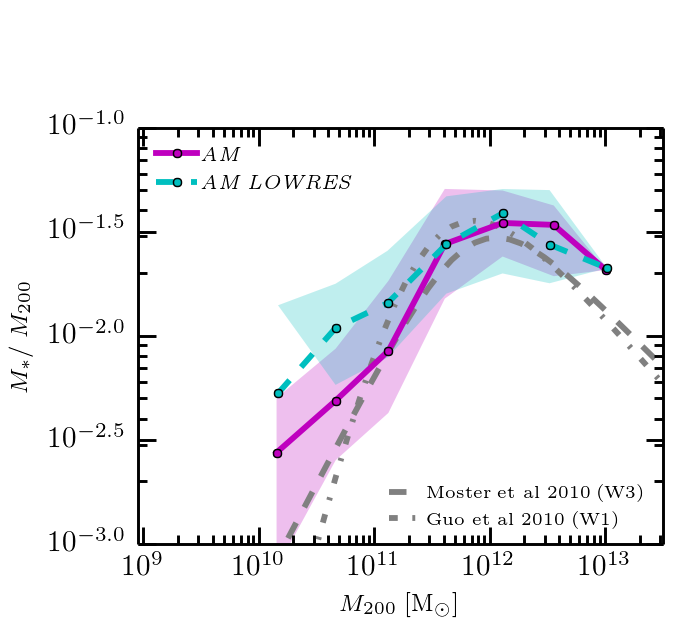}
&
\includegraphics[width=\columnwidth]{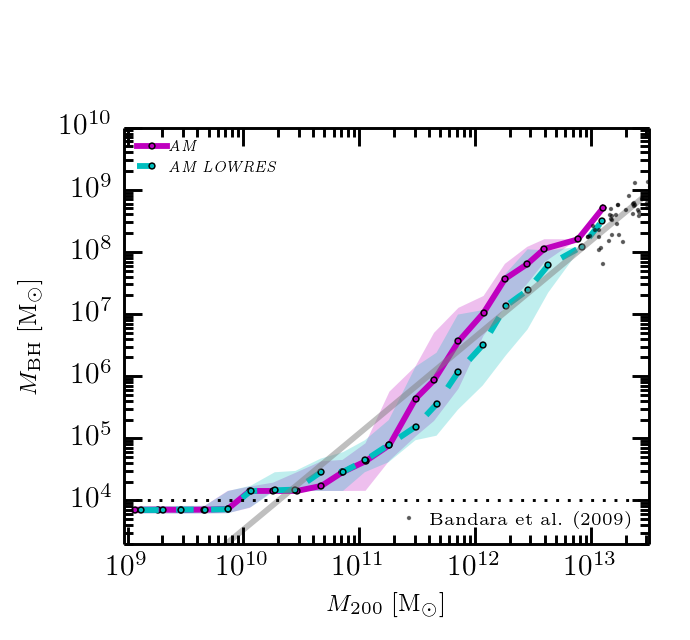}
\\
\end{tabular} 
\caption{The right panel shows the correlation of stellar mass fraction with $\mcrit$, while  
 the left panel shows the $\mbh$--$\mcrit$ relation. The simulation {\it AM} is shown as a purple solid line 
while  the simulation {\it AM LOWRES}, with lower resolution in mass,
is included as  cyan dotted line. Although there is some residual offset in BH masses in the right hand panel, the stellar
mass fractions (left hand panel) shows good convergence in haloes that are  well resolved in both simulations.}
\label{fig:resolutiontests} 
\end{figure*}

Fig.~ \ref{fig:resolutiontests} shows the stellar mass fraction--$\mcrit$ relation and  
the $\mbh$--$\mcrit$ relation for these simulations in 
purple and cyan for well resolved halos in each simulation. The stellar mass fractions 
have similar values of the medians and scatter. From this measure, the convergence seems good.
In the case of the $\mbh$--$\mcrit$ relation,  the medians take somewhat smaller values for the low-mass BHs 
in the simulation {\it AM LOWRES}, but the two resolutions show a large overlap in terms of the scatter 
and appear to converge at high and low masses. The convergence is nevertheless better than for the accretion disk particle scheme \citep{power2011} when it is applied at this resolution \citep{muldrew13}.

\section{The effect of changing effective viscosity parameter $ \Cvisc$ }
\label{appndx:variation_alphavisc}

\begin{figure*} 
\begin{tabular}{cc}
\includegraphics[width=\columnwidth]{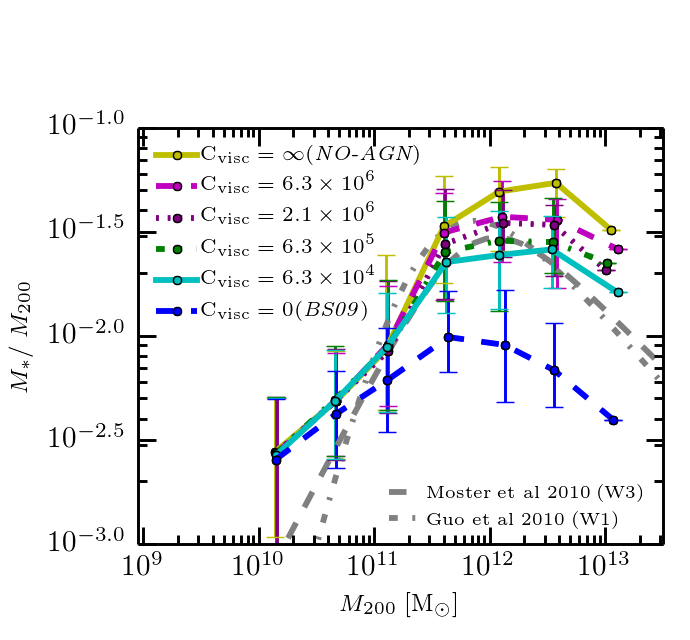}
&
\includegraphics[width=\columnwidth]{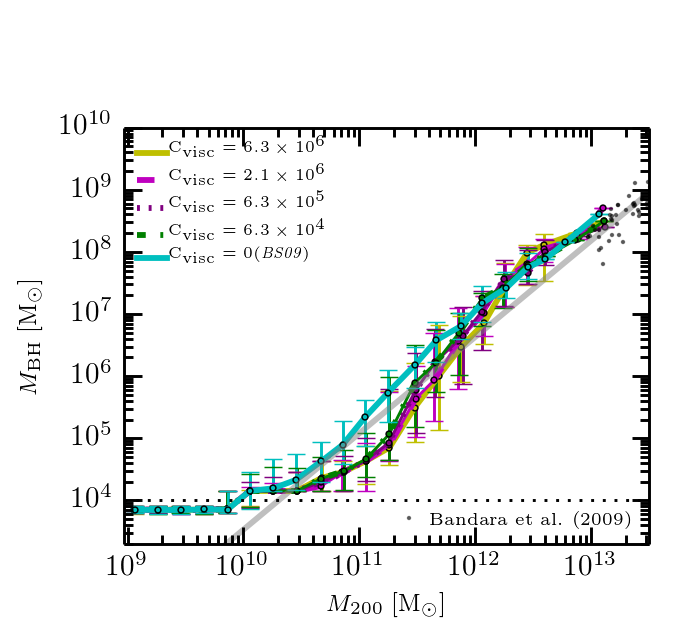}
\\
\end{tabular} 
\caption{The left panel shows the stellar mass fraction-$\mcrit$ relation and the right panel shows the $\mbh$--$\mcrit$ 
relation for simulations similar to {\it AM } ($\Cvisc=2.1 \times 10^6$), but with varying $\Cvisc$ as is
indicated in the legends. The simulations  {\it NO-AGN} ($\Cvisc=\infty$, yellow solid colour) and {\it BS09} 
    ($\Cvisc=0$, blue dashed colour) show extreme cases of the effects of $\Cvisc$. The smaller $\Cvisc$,the higher the suppression
   in the stellar mass fraction is above a critical mass halo. The critical  halo mass decreases slightly as  $\Cvisc$ decreases. 
The $\mbh$--$\mcrit$ relation is not affected above this critical mass, but as $\Cvisc$ decreases slightly less massive BHs 
are hosted by halos below this critical mass.}
\label{fig:varyingcvisc} 
\end{figure*}

  In this appendix, we show the effect of changing the value of the effective viscosity parameter on the 
  stellar mass fraction in the simulated galaxy population and on the $\mbh$--$\mcrit$ relation. We increase
  $\Cvisc$ by two orders of magnitude from $6.3 \times 10^4$ to $6.3 \times 10^6$ (our fiducial value is $2.1 \times 10^6$, purple line). The Fig.~
  \ref{fig:varyingcvisc} shows the stellar mass fraction and the $\mbh$--$\mcrit$ relation when $\Cvisc$
  takes these values. We have included  the simulations {\it NO-AGN} ($\Cvisc=\infty$, yellow colour) and {\it BS09} 
   ($\Cvisc=0$, blue colour) as extreme cases of $\Cvisc$.  
 
 Decreasing $\Cvisc$ (corresponding to a higher disk viscosity or thicker disk) results in the haloes being regulated by BH accretion at slightly lower stellar masses and  as a result the stellar mass fraction (left panel) breaks at a critical halo mass, however, how sharp the turnover in the stellar mass fraction is depends on $Cvisc$. The extreme case of this effect is shown in 
  the simulation {\it BS09} ($\Cvisc= 0$) where the stellar mass fraction lies below the abundance matching results. For less extreme values, the dependence on $\Cvisc$ is weak, re-enforcing the idea that the break in the mass fraction is set by the impact of periods
of AGN heating (and hence by the halo's cooling and dynamical time scales) rather than on the details of the BH accretion model. 

  Looking at the  $\mbh$--$\mcrit$ relation 
  (right panel), for haloes above mass $10^{12}\msun$ variations in $\Cvisc$ have little effect: in this regime, BHs are able to  self-regulate their growth regardless of the details of the accretion model.  Halos below this critical mass  tend to host less
  massive BHs, although the dependence on value of  $\Cvisc$ is again weak. 
  
 Hence, the main results are relatively insensitive to variations of $\Cvisc$ regarding  to  the critical halo mass in which the break of stellar mass fraction occurs, reproducing a turnover for halos with mass larger than the critical mass. The sharpness  of the turnover depends on the choice of $\Cvisc$. Large changes in $\Cvisc$ are required to have significant impact because the suppression by angular momentum is only effective when $(\Cvisc)^{1/3} \Vphi> \cs$ (see. eq.\ref{eq:newbhacr}). Although the changes in $\Cvisc$ affect the sharpness of the turnover in the mass fraction, this allows adjustment of the location of the break in the stellar mass fraction seen in cosmological volume simulations.

\bsp

\label{lastpage}

\end{document}